\documentclass[showpacs,amsmath,amssymb,aps,prd,lengthcheck]{revtex4}

\usepackage{graphicx}
\usepackage{dcolumn}
\usepackage{bm}
\usepackage{hyperref}
\usepackage{amssymb}
\usepackage{graphicx}
\usepackage{subfigure}

\begin{document}


\title{Maximum gravitational-wave energy emissible in magnetar flares}

\author{Alessandra Corsi}

\affiliation{
LIGO Laboratory,
California Institute of Technology,
MS 100-36,
Pasadena, California 91125, USA}

\author{Benjamin J. Owen}

\affiliation{
Center for Gravitational Wave Physics,
Institute for Gravitation and the Cosmos,
Department of Physics,
The Pennsylvania State University, University Park, Pennsylvania 16802, USA
}
\affiliation{
Max Planck Institut f\"ur Gravitationsphysik (Albert Einstein Institut),
Callinstr.\ 38, 30167 Hannover, Germany
}

\date{February 16, 2011}

\pacs{
04.30.Db, 
04.30.Tv, 
97.60.Jd, 
95.85.Sz  
}

\begin{abstract}

Recent searches of gravitational-wave (GW) data raise the question of what
maximum GW energies could be emitted during gamma-ray flares of highly
magnetized neutron stars (magnetars).
The highest energies ($\sim 10^{49}$~erg) predicted so far come from a model
 [K. Ioka, Mon.\ Not.\ Roy.\ Astron.\ Soc. {\bf 327}, 639 (2001)] in
which the internal magnetic field of a magnetar experiences a global
reconfiguration, changing the hydromagnetic equilibrium structure of the star
and tapping the gravitational potential energy without changing the magnetic
potential energy.
The largest energies in this model assume very special conditions, including a
large change in moment of inertia (which was observed in at most one flare), a
very high internal magnetic field, and a very soft equation of state.
Here we show that energies of $10^{48}$--$10^{49}$~erg are possible under
more generic conditions by tapping the magnetic energy, and we note that
similar energies may also be available through cracking of exotic solid cores.
Current observational limits on gravitational waves from magnetar fundamental
modes are just reaching these energies and will beat them in the era of
advanced interferometers.

\end{abstract}

\maketitle

\section{Introduction}
\label{Sec:Intro}

\subsection{Motivation}

Recent years have seen the publication of several searches for
gravitational-wave (GW) bursts triggered by gamma-ray flares from soft gamma
repeaters (SGRs) and anomalous x-ray pulsars (AXPs), both of which are
believed to be highly magnetized neutron stars (magnetars).

The most sensitive searches are from the Laser Interferometer
Gravitational-wave Observatory (LIGO) and Virgo, targeting the 2004 giant
flare from SGR\,1806--20 as well as many smaller flares from up to six
magnetars \citep{Abbott:2007zzb, Abbott:2008gj, Abbott:2009zd, Abadie:2010wx}.
No GW signals were found, and thus the results are upper limits on the GW
energy emitted as low as $\sim 10^{48}$~erg for fundamental or $f$-modes
(frequencies above $10^3$~Hz) or $\sim 10^{45}$~erg for frequencies of
greatest LIGO and Virgo sensitivity ($\sim 10^2$~Hz) \citep{Abadie:2010wx}.
The best (lowest) energy limits on the 2004 giant flare (which emitted
$\sim 10^{46}$~erg in photons) were $\sim 10^{51}$~erg for $f$-modes and
$10^{48}$~erg at $10^2$~Hz \citep{Abbott:2008gj}.
Similar best energy limits on the 2009 ``ring'' event (which is now believed
to have been a giant flare emitting $10^{44}$--$10^{45}$~erg in photons) were
$\sim 10^{49}$~erg and $10^{46}$~erg \citep{Abadie:2010wx}.
In a few years, when LIGO and Virgo are upgraded to ``advanced
interferometer'' status, their noise amplitudes will improve by an order of
magnitude \citep{aLIGO, aVirgo} and thus energy sensitivities will improve by
two orders of magnitude.

Present upper limits and predicted sensitivities raise the question of
what maximum GW energies could possibly be radiated during magnetar flares.
In spite of its relevance for ongoing and rapidly improving searches for GWs
from magnetars, there has been relatively little work on this question.
The closely related question of what is the ratio of GW-emitted energy to
electromagnetically (EM) emitted energy is not addressed at all in the
literature, with searches therefore relying on possible correlations between
observables \citep{Kalmus2009}.
We do note that a high GW/EM energy ratio, which is relevant to current GW
observations, might be possible if most of the action takes place in the
interior of the star, as suggested by recent work of \citet{LanderJones2011}.
A high GW/EM energy ratio might also explain flares with high energy, but no
initial spike or pulsations (typical of giant flares), as observed in  
SGR\,1627--41 \citep{Mazets1999}.
But in this article we concern ourselves only with the maximum available
energies.
In the rest of the Introduction we discuss the two major models: the
crust-cracking model and the hydromagnetic deformation model.

\subsection{Crust cracking model}

The now-standard interpretation of SGR flares within the magnetar model of a
highly magnetized neutron star is that they involve the solid crust of the star
cracking as it is strained by twisting magnetic field lines, with the field
rearranging itself afterwards \citep{Duncan1992, ThompsonDuncan1995}.
This is supported by the good fit of SGR flare gamma-ray energy and waiting
time distributions to the universal power laws for brittle fracture
\citep[e.g.][]{Cheng1996, Palmer1999, Dubath2005, Perna:2011gt}.
Some of the energy of the cracking event should excite quasinormal modes of the
star. Indeed there is evidence from Quasi Periodic Oscillations (QPOs) in x-ray tails of giant flares
that shear modes or torsional modes of the solid crust are excited, possibly coupled
to magneto-hydrodynamic modes in the core \citep{Israel:2005av,
Strohmayer:2005ks, Levin:2006qd}.

We note that, even under the hypothesis that the flare originates in the magnetosphere 
\citep[see e.g.][]{Lyutikov:2005un}, the magnetospheric
reconfiguration exerts magnetic stress on the crust which can hydromagnetically
couple to modes in the core \citep{Levin:2006qd, vanHoven:2010gy,
Gabler:2010rp}.

In the above scenarios, the flare should excite to some extent the fundamental
or $f$-modes of the star, which radiate GW with damping times of $\sim$200~ms
\citep{Detweiler1983, Pacheco1998, Gualtieri2004}.
These timescales are shorter than other relevant ones, except
for the Alfv\'en-wave crossing time of the star, to which they are comparable. Therefore,
the $f$-modes are likely to radiate most of the energy they receive as GWs, even if other modes are
excited to higher energies by the event that causes the gamma-ray flare.
And, if much of the flare energy goes into exciting the $f$-modes, they might
emit GW energy exceeding the emitted EM energy.

The details of which modes are most excited and what are likely GW to EM emission energy
ratios are even more difficult to address than the total energy budget,
and have not yet been investigated in the literature.
Therefore we, like previous authors, restrict our attention to the total
energy budget of the largest SGR flares, which serves as an upper limit to the
GW energy emitted.

A natural estimate for the maximum GW energy radiated by the crust-cracking
mechanism is the maximum elastic deformation energy of the crust, which should
be at least comparable to largest gamma-ray energy emitted in a giant flare.
The EM energy emitted in the 2004 giant flare of SGR\,1806--20
\citep{Hurley:2005zs}, of order $10^{45}$--$10^{46}$~erg, was greater than
previous giant flare energies and hard to reconcile with the standard maximum
crust elastic energy of order $10^{44}$~erg \citep[e.g.][]{Blaes1989}.
The latter energy is proportional to the shear modulus of the solid part of the
star, and thus the 2004 giant flare energy could be explained by solid quark
matter.
With a shear modulus exceeding that of a neutron-star crust by 3--4 orders of
magnitude \citep{Horvath:2005ta, Owen:2005fn, Xu:2006mp}, energies of order
$10^{47}$--$10^{48}$~erg would become available.

The maximum crust elastic energy is also proportional to the square of the
breaking strain of the material which, until recently, was usually assumed to
be at most $10^{-2}$, comparable to the best terrestrial alloys.
Molecular dynamics simulations by \citet{Horowitz:2009ya}, though strictly
applicable only to the outer crust at densities below neutron drip, indicate
that the breaking strain of dense solid matter can reach $10^{-1}$ as defects,
domain walls, etc.\ are crushed away by the intense pressure.
Using the above scaling, this brings the maximum elastic energy (and thus GW
energy) up to $10^{46}$~erg for a normal neutron star, reconciling it with the
EM energy emitted in the 2004 giant flare.

We note, apparently for the first time in the literature, that even higher energies are possible from the cracking mechanism if
the neutron star or at least its core is made of a solid form of quark matter,
and the breaking strain of that matter is of order $10^{-1}$.
In fact, \citet{Horowitz:2009ya} restrict their simulations to the low-density outer
layers of a normal nuclear matter crust, and do not speculate on the physics of
exotic phases with or without strong magnetic fields. However, the crushing of
defects under intense pressure which is responsible for a high breaking strain
seems to be robust physics.

From the above mentioned scalings and from shear modulus calculations in
the literature, we infer that, if the high breaking strain of \citet{Horowitz:2009ya} is
generic, GW energies of order $10^{48}$~erg are possible for mixed
baryon-meson or baryon-quark phases \citep{Owen:2005fn}, and energies of order
$10^{49}$--$10^{50}$~erg are possible for solid quark phases \citep{Xu2003,
Mannarelli2007}.
A more careful estimate of the former is forthcoming \cite{exotic}.

\subsection{Hydromagnetic deformation model}

The highest GW energies previously obtained in the literature and noted in the
$f$-mode searches \citep{Abbott:2008gj, Abbott:2009zd, Abadie:2010wx} come
from a model by \citet{Ioka2001} based on magnetic deformations of the star's
hydrostatic equilibrium.
These can be $10^{48}$--$10^{49}$~erg, comparable to the latest upper limits
on GW emission from $f$-modes \cite{Abadie:2010wx}.

It may seem surprising that magnetar flares could be good candidates for GW
detection given that supernovae, with a total EM energy emitted orders of
magnitude above that emitted in giant flares, are difficult targets for GW
searches even with improved instruments and algorithms \citep{Ott:2008wt}.
Although the EM energy release in supernovae is large, the bulk motion of
matter which generates GWs mainly involves material at densities lower than
nuclear density and features relatively little quadrupolar motion.
A rearrangement of the interior of a neutron star, on the other hand, involves
matter at supernuclear densities, and the magnetic dipole couples directly to
the mass quadrupole through the magnetic pressure.

While most neutron stars have external magnetic dipole fields less than $\sim
10^{12}$--$10^{13}$~G, there is growing evidence for the existence of
super-magnetized neutron stars with fields of $\sim 10^{14}-10^{15}$~G
\citep{ThompsonDuncan2001, Thompson2002}.
Larger magnetic fields of $\sim 10^{16}$ G may be generated by the helical
dynamo inside a newborn neutron star \citep{Duncan1992,ThompsonDuncan1993}, and
even the maximum field strength allowed by the virial theorem ($10^{18}$\,G)
could be achieved if the central engines of gamma-ray bursts are magnetars
\citep{Usov1992,Klu1998,Wheeler2000}.
Internal fields of order $10^{16}$~G are also suggested by lifetime energetics
and cooling models and observations of persistent x-ray emission
\cite{KaminkerEtAl2006}.
An internal field of $10^{16}$~G puts the ratio of magnetic potential energy
($\sim 10^{49}$~erg) to gravitational potential energy ($\sim 10^{53}$~erg) at
$10^{-4}$.

\citet{Ioka2001} noted that an increase in the spin period of SGR~1900+14 by a
fraction $10^{-4}$ over an 80-day interval including its 1998 giant flare
could have been produced by a sudden $10^{-4}$ fractional increase in the
moment of inertia at the time of the flare, which in turn could have been
related to a reconfiguration of a toroidal internal magnetic field.
The internal magnetic field is believed to be mainly toroidal due to dynamo action
in the first few seconds of the star's life \citep{Duncan1992,ThompsonDuncan1993,ThompsonDuncan2001}.
A mainly toroidal field makes the star prolate, leading to an increase in the
moment of inertia when energy is released.

With some simplifying assumptions described below, \citet{Ioka2001} found a set
of stellar equilibria with discrete energies and moments of inertia.
For his most realistic equation of state (EOS), an $n=1$ polytrope (see below),
\citet{Ioka2001} found states separated by $\Delta {\cal I}/{\cal I}=10^{-4}$ in moment of inertia and
$10^{45}$~erg in energy, roughly the observed EM energy of the 1998
giant flare.
In order to have energy differences between equilibria close to $10^{45}$~erg, 
being $\Delta {\cal I}/{\cal I} \sim \delta$, with $\delta$ being the magnetic/gravitational energy ratio,
\citet{Ioka2001} chose flare models (i.e.\ jumps between equilibria)
which kept the magnetic energy constant.
This made the overall energy release second order in $\delta=10^{-4}$:
$10^{53}$~erg $\times$ $(10^{-4})^2 = 10^{45}$~erg.
\citet{Ioka2001} also gave energies for very soft EOS (high
polytropic index) and high internal magnetic field (more than $10^{17}$~G)
which were up to nearly $10^{49}$~erg, comparable to recent observational upper
limits on GW emission.

Motivated by these high predicted energies, we re-examine the model of
\citet{Ioka2001} with an eye toward exploring its broader applicability and
robustness, and we push it to find under what conditions the highest GW
energies are possible.

\subsection{Outline}

First to be addressed in generalizing the model by \citet{Ioka2001} are several simplifying assumptions such as Newtonian
gravity, symmetry, lack of superconductivity, and polytropic equation of
state.
In Section~\ref{Sec:Justification}, we argue that the message to be drawn from the
more detailed works appearing in the literature during the years since 2001
is that the physically simplified model of \citet{Ioka2001} well serves our
present goal of estimating the order of magnitude of energy available.

In Section~\ref{Sec:Math} we describe Ioka's choice of magnetic field and the rest
of the mathematical formalism (the first-order part of his calculation).

In Section~\ref{Sec:Results} we show that the model by \citet{Ioka2001} has applicability beyond the
1998 giant flare. The biggest concern with such a model is, in fact, that it 
was built to explain the putative $10^{-4}$ change in spin period after the 1998 giant flare of SGR~1900+14.
However, such changes are not observed associated with most flares; and indeed
the data for the 1998 flare itself could be interpreted in other ways such as
timing noise \citep{WoodsEtAl2007} or change of the external dipole field.
We qualitatively discuss the broader possibilities for jumps between equilibria, and 
we give quantitative results for a particular family of jumps which 
tends to produce larger energies with smaller moment of inertia changes.

In Section~\ref{Sec:Discussion} we summarize the results of our explorations 
and discuss their consequences for current and future GW
searches by LIGO and Virgo.


\section{Physical assumptions and justification}
\label{Sec:Justification}

In this Section we address the accuracy of a number of simplifying assumptions used in the
analysis by \citet{Ioka2001}, which we also adopt here.
Most of them have been investigated further in recent years in the context of
continuous GW emission from newborn magnetars.

\subsection{Perturbative approach}

Our first assumption is that the effect of the magnetic field on stellar
equilibria is much greater than that of rotation, and much less than that of
gravity.
This is straightforward to check as in \citet{Ioka2001}:
The magnetic field is a perturbative effect on the hydrostatic equilibrium of
the star if the typical magnetic field strength satisfies $H \ll
10^{18}(R/10^{6}{\rm cm})^{-4}(M/M_{\odot})^{2}$~G, which it does even for the
fields $H\sim10^{16}$~G predicted inside magnetars.
The internal magnetic field induces a deformation which dominates the
rotational one when $H \gg 10^{14}(P/1 {\rm~s})^{-1}$~G, where $P$ is the spin
period.
For SGRs, $P$ is of the order of 5--10~s \citep[see e.g.][]{Mereghetti2008},
and thus rotation can be neglected. 

A recent calculation \citep{Lander2009} including rotation and nonlinear
magnetic equilibrium confirms that these are negligible effects for the systems
considered here.
Neglecting these effects allows adopting a formalism similar to that developed
by \citet{Chandrasekhar1933} and \citet{Chandrasekhar1962} for slowly rotating
polytropes, in which the perturbation parameter is the ratio of the rotational
to gravitational energy. In cases where the magnetic field is the sole
perturbation, the perturbation parameter becomes the ratio of magnetic to
gravitational potential energy \citep{Chandrasekhar1953, Monaghan1965,
Monaghan1966a, Monaghan1966b, Roxburgh1966, Trehan1971, Trehan1972}.

Like almost all other authors, we neglect the effect of stable stratification
(non-barotropic composition gradients) on the hydromagnetic equilibrium,
although this may come into play on longer timescales, such as the cooling
timescale \cite{Reisenegger2009}.

\subsection{Relativistic gravity corrections}

The effects of relativistic gravity have also been investigated in recent
years.

In Newtonian analyses such as \citet{Ioka2001}, the magnetic stress of a
toroidal field tends to make the star prolate, working like a rubber belt
tightening up the equator of the star; and the analysis by \citet{Ioka2004}
confirms the validity of this picture in relativistic stars.

More specifically, \citet{Ioka2003} and \citet{Ioka2004} extended the results of \citet{Ioka2001}
to relativistic gravity (for an $n=1$ polytrope).
They obtained stationary axisymmetric configurations of magnetized stars in the framework of general relativistic ideal magnetohydrodynamics, incorporating a toroidal magnetic field and
meridional flow, in addition to a poloidal magnetic field. As in \citet{Ioka2001}, \citet{Ioka2004} worked under the hypothesis of axisymmetry; boundary conditions so as to have the magnetic field vanishing at the stellar surface; and magnetic field weak compared to gravity, so that it can be treated as a small perturbation on an already-known non-magnetized, non-rotating configuration.
They found an eigenvalue problem with energies separated by nearly
$10^{48}$~erg for internal fields of order $10^{16}$~G.
(This is obtained from their Table~2, second group of rows---the first is
unstable---multiplying column~3 by column~9 and keeping in mind that their
$\mathcal{R}_M$ is slightly greater than our $\delta$.)
These energies are nearly two orders of magnitude greater than the $n=1$ jumps
from Fig.~3 of \citet{Ioka2001}, more comparable to the jumps for the
marginally stable $n=2.5$ EOS.
Relativity increases the central condensation of the star compared to
Newtonian gravity and thus is expected to give numbers comparable to softer
(higher-$n$) EOS.
Therefore our Newtonian energy estimates for $n=1$ in fact should be somewhat
conservative.

Other relativistic analyses \citep[e.g.\ ][]{Colaiuda2008, Ciolfi2009,
Ciolfi2010} change even more features of the analysis of \citet{Ioka2001}, as
we discuss in the next Sections.

\subsection{Boundary condition}
\label{boundary}

More important are the interlinked issues of magnetic field configurations,
especially the toroidal-to-poloidal ratio and boundary conditions at the
surface of the star, and the EOS.

The discrete energy spectrum at the heart of the model by \citet{Ioka2001}, is due
to the boundary condition imposed on the magnetic field at the surface.
This may seem to be a very specialized condition, but we argue that it is more
generally applicable.

\citet{Ioka2001} takes the toroidal part of the field to vanish at the stellar
surface, which has the effect of forcing surface and magnetospheric currents
to vanish.
\citet{Ioka2001} also assumes a field configuration with a fixed toroidal to poloidal ratio,
so that the poloidal field vanishes at the surface too.
Both assumptions are common in the literature.
The latter is an issue since the observed spin-downs of magnetars usually
imply external dipole fields of $10^{14}$--$10^{15}$~G just outside the
surface.
However this is small compared to the internal field, and there is now
observational evidence for a magnetar with a large internal field and even
smaller (less than $10^{13}$~G) external dipole field \cite{ReaEtAl2010}.

Invoking surface currents \citep{Colaiuda2008} can set the toroidal field
discontinuously to zero just outside the star (compared to a finite value just
inside) while letting the internal poloidal field be matched to an external
dipole.
However, there is little to indicate what the surface currents on a neutron star
should be, and thus they are neglected in most studies
\citep{Ioka2004,Haskell2008,Ciolfi2009,Ciolfi2010,Lander2009}.
A barotropic EOS (dependent only on pressure) with density going to zero at the
surface also forces the magnetic field to go to zero at the surface in ideal
magneto-hydrodynamics (MHD) \citep{Haskell2008}.
However, magnetic diffusivity due to resistance can be invoked to get around
that problem \citep{Braithwaite2008}:
Neutron stars are not perfect conductors, and in moving from the superfluid
interior to the crust and magnetosphere, the resistivity of the medium should
increase and hence the boundary conditions should be adapted to reflect this
behavior.

At any rate, if the internal field is matched to a much smaller external field
the result should not differ greatly from matching to zero external field.
Spin-down observations argue that the external dipole field does not change
greatly even in giant flares \citep{Ioka2001, WoodsEtAl2007}.
Matching to any fixed external field will still result in discrete eigenvalues,
so the mechanism should not be qualitatively changed and one would estimate is
quantitatively changed by of order $H_\mathrm{ext}/H_\mathrm{int}$ or of order
10\% for the scenario envisioned here.

The conclusion we draw from these works is that, while the no-external-field
boundary condition is obviously a specialized simplification, the crucial
property of discrete eigenvalues has greater generality.

\subsection{Toroidal-to-poloidal field ratio}
There has been much work on the toroidal-to-poloidal field ratio as well.

Recently \citet{Lander2009} studied the various stationary, axisymmetric equilibrium solutions for Newtonian fluid stars in perfect MHD, showing that the full equations of MHD reduce under these limits to two general cases: a mixed-field case (which includes purely poloidal fields as a special case) and purely toroidal fields. 

In the mixed-field case, differently from the boundary condition of zero
exterior field set by e.g.\ \citet{Ioka2001} and \citet{Haskell2008}, the
toroidal field component is set to vanish outside the star (i.e., no currents
exist on or outside the neutron star's surface), while the poloidal
field is matched through the stellar surface to an external dipole vanishing at
infinity. \citet{Lander2009} find that the equilibrium configurations are poloidal-dominated.

The boundary condition being the main difference between \citet{Haskell2008}
and \citet{Lander2009}, the latter authors conjecture that matching to an
outside dipole field favors poloidal-dominated fields and oblate stars, while a
vanishing magnetic field on the surface favors toroidal-dominated fields and
prolate stars. As \citet{Lander2009} have emphasized, for a real neutron star the resistivity
of the outer layers could resemble a boundary condition intermediate between
the two cases.

Various studies dedicated to finding mixed field equilibrium configurations
with specific boundary conditions \citep{Tomimura2005, Yoshida2006a,
Yoshida2006b, Ciolfi2009, Ciolfi2010, Ioka2004, Colaiuda2008, Haskell2008, Lander2009} resulted in different poloidal-to-toroidal field ratios. 
The configurations obtained with the boundary condition set by
\citet{Lander2009} all have no more than $7\%$ of the magnetic energy stored
in the toroidal field component.
Ciolfi et al.\ \cite{Ciolfi2009, Ciolfi2010} also found that in their
configurations, although the amplitudes of both the poloidal and toroidal
fields are of the
same order of magnitude, and the toroidal field in the interior can be larger
than the poloidal field at the surface, the contribution of the toroidal field
to the total magnetic energy is $\lesssim 10\%$, because this field is non
vanishing only in a finite region of the star.
On the other hand, by setting the magnetic field to vanish outside the star,
\citet{Ioka2001} (whose results agree with \citet{Haskell2008} for the case of
a $n=1$ polytrope) obtains equilibrium configurations where up to $\sim 96\%$ of
the magnetic energy is stored in the toroidal component (see line 10 in Table \ref{Tab1}). 

An interesting point is how these results compare with those from studies aimed at evaluating the actual stability of magnetic equilibria in stars. These have shown that a stellar magnetic field in stable equilibrium must
contain both poloidal (meridional) and toroidal (azimuthal) components, since
both are unstable on their own \citep{Prendergast1956, Markey1973, Tayler1973,
Wright1973, Markey1974, Spruit1998, Braithwaite2006a, Braithwaite2006b,
Braithwaite2006c, Bonanno2008, LanderJones2011}. Stars with purely poloidal magnetic fields suffer from a hydromagnetic
instability, while the instabilities are
suppressed if the toroidal magnetic fields in the star have comparable
strength with the poloidal fields \citep{Yoshida2006a}. 

Numerical evolutions by \citet{Braithwaite2008} give indications that the toroidal field component should store 20--90\% of the total magnetic energy
in order for the neutron star to be stable. Via MHD simulations, \citet{Braithwaite2004}
have found that purely poloidal magnetic fields in stars decay completely within a
few Alfv\'en timescales, while ``twisted-torus'' poloidal-toroidal mixed configurations can survive for times much longer than the Alfv\'en time. These configurations are roughly axisymmetric; 
the poloidal field extends throughout the entire star and to the exterior, while the
toroidal field is confined in a torus-shaped region inside the star, where the
field lines are closed \footnote{This particular field geometry is obtained assuming vacuum outside the star,
i.e.\ electric currents are forbidden outside and the magnetic field can only
be poloidal. Different solutions including a magnetosphere may be possible, where the
toroidal field could also extend to the external region, leading to a twisted
magnetosphere \citep[see e.g.][]{Pavan2009}.}. 

In this paper, we follow \citet{Ioka2001} and consider equilibrium states where the contribution of the toroidal field energy is in between $\sim 65\%$ and $\sim 96\%$ of the total magnetic energy in the star (Table\,\ref{Tab1}).

We finally note that most of the above mentioned works have considered normal fluid stars, although neutron stars are
believed to become superconducting superfluids over much of their volume
shortly after birth. The latter case is much more complicated to treat, but see \citet{Akgun2008}
for a recent careful calculation indicating that mostly toroidal fields may be
stable in this case too.

\subsection{Equation of state}
A final issue is the dependence on the EOS.

\citet{Kiuchi2008a} have considered Newtonian magnetized stars
with four kinds of realistic EOSs (SLy by \citep{Douchin2001}; FPS by
\citep{Pandharipande1989}; Shen by \citep{Shen1998}; and LS by \citep{Lattimer1991}). For the
non-rotating sequences, they found that there exist nearly toroidal
field configurations, irrespective of the EOSs. The magnetic
energy stored in the stars increases with the degree of deformation
being larger.

More recently, \citet{Kiuchi2009} have investigated equilibrium sequences of 
relativistic stars containing purely toroidal magnetic fields, with the same
four kinds of realistic EOSs. In the non-rotating case, it is found that for a SLy EOS, the toroidal magnetic field peaking in the vicinity of the equatorial plane acts through the Lorentz forces to pinch the matter around the magnetic axis, making the stellar shape prolate. Indeed, the toroidal magnetic field lines behave like a rubber belt that is wrapped around the waist of the star. This gross property is common to the other realistic EOSs \citep{Kiuchi2009}. 

For equal values of the central density, the profiles of stars with SLy and FPS EOSs are quite similar, while the density distribution of the star with Shen EOS is less prolate than SLy and FPS EOSs. The concentration of the magnetic field to the stellar center for Shen EOS is weaker than that for SLy or FPS EOS, the matter pressure stays relatively large up to the stellar surface, and the regions in which the magnetic pressure is dominant over the matter pressure appear rather in the outer regions. This implies that the magnetic fields for Shen EOS are effectively less fastening to pinch the 
matter around the magnetic axis than those for SLy or FPS EOS \citep{Kiuchi2009}. 

For LS EOS, the regions in which the ratio of the magnetic pressure
to the matter pressure is large also exist near the stellar surface.
However the density distribution is found to become similar to that for SLy and FPS EOSs 
because the pressure ratio is sufficiently higher than that for Shen EOS \citep{Kiuchi2009}.

In conclusion, since relativistic corrections to gravity, boundary conditions
and EOS do not seem to prevent the existence of prolate states of equilibrium
sustained by strong toroidal fields, the simplified treatment by
\citet{Ioka2001} is valid for the purpose of estimating the order of magnitude
of the maximum GW energy that may be released in jumps between equilibria.


\section{Mathematical formalism}
\label{Sec:Math}

In this Section we review the mathematical formalism for the equilibria of 
magnetized polytropes and the particular choice of magnetic field configuration used by
\citet{Ioka2001}.
We basically follow his results, simplifying the presentation so as to
concentrate only on the fundamental passages relevant for our work \citep[but
more details can be found in][]{LIGOnote}, while giving all the 
necessary elements to understand the underlying physics.
For an immediate comparison of our results with the ones by \citet{Ioka2001},
we also keep his notation.

\subsection{Equilibrium Equations}

Consider a non-relativistic, one-component perfectly conducting fluid in
hydrostatic equilibrium, with a magnetic field and vanishing net charge 
(as typical for astrophysical fluids or plasmas). The equations governing the 
equilibrium are
\begin{eqnarray}
\label{eq1}
-\nabla p+\rho \nabla\Phi+\frac{1}{4\pi}(\nabla\times\vec{H})\times\vec{H} &=& 0,
\\
\label{eq2}
\nabla^{2}\Phi &=& -4\pi G \rho,
\\
\label{eq3}
\nabla\cdot\vec{H}&=&0	,
\end{eqnarray}
where $\rho$ is the mass density, $p$ is the pressure, and $\Phi$ is the 
gravitational potential. The first is the Euler equation for a non-rotating
magnetized conducting fluid.
The second is Poisson's equation and the last is one of 
Maxwell's equations. We further assume a polytropic EOS 
\citep{Chandrasekhar1967}
	\begin{equation}
	p=K \rho^{1+1/n},
	\label{press}
	\end{equation}
and use this equation and the corresponding length scale
\begin{equation}
	\alpha=\left[\frac{(n+1)K \rho^{1/n-1}_c}{4\pi G}\right]^{1/2}
	\label{alpha}
\end{equation}
in terms of the central density $\rho_c$,
to convert to dimensionless variables $\Theta$, $\xi$,  $\vec{h}$, $\phi$,
defined as follows:
	\begin{eqnarray}
	\rho&=&\rho_c\Theta^n,
	\label{newdefrho}
\\
	r&=&\alpha\xi,
	\label{newdefr}
\\
	\vec{H}&=&(4\pi G \delta)^{1/2} \rho_c \alpha \vec{h},
	\label{newdef}
\\
	\Phi&=&4\pi G \alpha^{2} \rho_c \phi  .
	\end{eqnarray}
Here $\delta$ is the ratio of magnetic to gravitational potential energy in
physical units.
In the dimensionless variables Eqs.~(\ref{eq1})--(\ref{eq3}) read \citep{Ioka2001}:
\begin{eqnarray}
\label{eq1new}
-\nabla \Theta + \nabla\phi+\frac{\delta}{4\pi \Theta^{n}}(\nabla\times\vec{h})\times\vec{h}&=&0,
\\
\label{eq2new}\nabla^{2}\phi&=&-\Theta^{n},
\\
\label{eq3new}\nabla\cdot\vec{h}&=&0	.
\end{eqnarray}

In the case of axisymmetry, $\vec{h}$ can be conveniently expressed in terms of
two scalar functions $P(\xi,\theta)$ and $T(\xi,\theta)$ as \citep{Woltjer1959}
\begin{eqnarray}
\label{decomp_H}
	\nonumber \frac{\vec{h}}{(4\pi)^{1/2}}=-\frac{1}{\widetilde{\omega}}\frac{\partial (\widetilde{\omega}^{2} P )}{\partial z}\hat{e}_{\widetilde{\omega}}+\widetilde{\omega} T \hat{e}_\varphi+\frac{1}{\widetilde{\omega}}\frac{\partial (\widetilde{\omega}^{2} P)}{\partial\widetilde{\omega}}\hat{e}_z=\\=\nabla\times (\widetilde{\omega} P \hat{e}_{\varphi}) +\widetilde{\omega} T \hat{e}_\varphi
\end{eqnarray}
where $\hat{e}_{\widetilde{\omega}}$, $\hat{e}_\varphi$, $\hat{e}_z$  are a 
unit vectors in the $\widetilde{\omega}$, $\varphi$, $z$ directions, and 
$(\widetilde{\omega},\varphi,z)$ are cylindrical coordinates, related to the 
spherical ones $(\xi,\theta,\varphi)$ by $\widetilde{\omega}=\xi\sin\theta$ 
and $z=\xi\cos\theta$.
Eq.~(\ref{eq1new}) implies that:
\begin{equation}
	\nabla\times\left(\frac{(\nabla\times\vec{h})\times\vec{h}}{4\pi\Theta^{n}}\right)=\nabla\times\vec{\cal L}=0,
	\label{calL}
\end{equation}
which is satisfied if $\vec{\cal L}$ is the gradient of a scalar function. It 
can be shown that for the case of an axisymmetric magnetic field one has
\begin{equation}
	\vec{\cal L}=\nabla N_P(\omega^2 P)
	\label{LNp}
\end{equation}
if the following relations hold \citep[see e.g.][]{Woltjer1959}:
\begin{eqnarray}
	\frac{\Delta_5 P}{\Theta^{n}}&=&-\frac{d N_P(\tau)}{d \tau}-\frac{T}{\Theta^{n}}~\frac{d N_T(\tau)}{d \tau},
	\label{delta5}
\\
	\widetilde{\omega}^{2}T&=&N_T(\tau)\label{omegaT},
\end{eqnarray}
with the five-dimensional Laplacian
\begin{equation}
	\Delta_5=\frac{\partial^{2}}{\partial z^{2}}+\frac{3}{\widetilde{\omega}}\frac{\partial }{\partial \widetilde{\omega}}+\frac{\partial^{2}}{\partial \widetilde{\omega}^{2}}.
\end{equation}
Here $N_T(\tau)$ and $N_P(\tau)$ are arbitrary functions of their argument 
$\tau=\widetilde{\omega}^{2} P$. Assigning to such functions a specific form, 
the corresponding $P(\widetilde{\omega},\theta)$ and 
$T(\widetilde{\omega},\theta)$ are found (and thus the magnetic field 
configuration is fixed) by solving Eqs.~(\ref{delta5})--(\ref{omegaT}) with 
appropriate boundary conditions. Once the magnetic field configuration is 
specified, the equilibrium density of the magnetized polytrope can be found 
solving (with appropriate boundary conditions) the modified Lane-Emden
equation \citep{Ioka2001}
\begin{equation}
	\nabla^{2} \Theta = -\Theta^{n}+\delta\nabla^{2}N_P(\widetilde{\omega}^{2} P),
	\label{finaltheta}	
\end{equation}
which is obtained by combining Eq.~(\ref{LNp}) with
Eqs.~(\ref{eq1new})--(\ref{eq2new}).


\subsection{Perturbative approach}

We assume that the solutions of Eqs.~(\ref{delta5}-\ref{omegaT}), 
(\ref{finaltheta}) have the following form \footnote{Note that according to Eq. 
(\ref{newdef}), the definition of $\vec{h}$ already contains a factor of 
$\delta^{1/2}$, so to obtain the poloidal and toroidal field components of 
$\vec{H}$ up to first order in $\delta$, it is sufficient to expand $P$ 
and $T$ in Eq. (\ref{decomp_H}) up to the zeroth order.}:
	\begin{eqnarray}
	P(\xi,\theta)&=&P_0(\xi,\theta)+{\cal O}(\delta),\label{devP}
\\
	T(\xi,\theta)&=&T_0(\xi,\theta)+{\cal O}(\delta),\label{devT}
\\
	\Theta(\xi,\theta)&=&\Theta_{0}(\xi)+\delta\Theta_{1}(\xi,\theta)+{\cal O}(\delta^2).\label{devtheta}
	\end{eqnarray}
Substituting into Eqs.~(\ref{delta5})--(\ref{omegaT}), one gets
\citep{Ioka2001}
\begin{eqnarray}
	\frac{\Delta_5 P_0}{\Theta^{n}_0}&=&-\frac{d N_P(\tau_0)}{d \tau}-\frac{T_0}{\Theta^{n}_0}~\frac{d N_T(\tau_0)}{d \tau},
	\label{delta50}
\\
	\widetilde{\omega}^{2}T_0&=&N_T(\tau_0),\label{omegaT0}
\end{eqnarray}
where $\tau_0=\widetilde{\omega}^{2} P_0$. Next, suppose that a particular 
choice for the magnetic field configuration is made by specifying the 
functions $N_P(\tau)$ and $N_T(\tau)$ and assigning boundary conditions 
for the magnetic field. Then by solving Eqs.~(\ref{delta50})--(\ref{omegaT0}), 
$P_0(\xi,\theta)$ and $T_0(\xi,\theta)$ are found. Further, performing a 
Legendre expansion, we can write:
\begin{equation}
	\label{leg3} N_P(\tau_0)=N_p(\widetilde{\omega}^{2} P_0(\xi,\theta))=\sum^{\infty}_{m=0} \Psi_m(\xi)P_m(\cos\theta)
\end{equation}
where $P_m(\cos\theta)$ denotes the Legendre polynomial of order $m$, and the 
coefficients $\Psi_m(\xi)$ are known once $P_0(\xi,\theta)$ is. To find the 
equilibrium configuration of the corresponding magnetized polytrope, one can 
then proceed as follows. We expand in Legendre polynomials the perturbed star 
density,
\begin{equation}
	\label{leg1}\Theta_1(\xi,\theta)=\sum^{\infty}_{m=0}\psi_m(\xi)P_m(\cos\theta),
	\end{equation}
where the coefficients $\psi_m$ are to be found. It is possible to show that 
Eqs.~(\ref{delta5})--(\ref{omegaT}), (\ref{finaltheta}) imply \citep{Ioka2001}
	\begin{eqnarray}
	{\cal D}_0 \Theta_0(\xi)&=&-\Theta^{n}_0(\xi),\label{ausild0}
	\\
	{\cal D}_m(\psi_m(\xi)-\Psi_m(\xi))&=&- n \Theta_0^{(n-1)}(\xi) \psi_m(\xi),\label{ausildm}
  \end{eqnarray}
using the dimensionless radial Laplacian
\begin{equation}
	{\cal D}_m=\left[\frac{1}{\xi^{2}}\frac{d }{d  \xi}\left(\xi^{2} \frac{d  }{d  \xi}\right)- \frac{m (m+1)}{\xi^{2}} \right].\label{dm}
\end{equation}

Equations~(\ref{ausild0})--(\ref{ausildm}) are to be solved by imposing the
boundary conditions
\begin{eqnarray}
	\Theta_0(0)=1, & \Theta'_0(0)=0 ,\label{boundfint}\\
	\psi_m(0)=0, & \psi'_m(0)=0, \label{boundfinpsi}
\end{eqnarray}
which assure that, to first order in $\delta$, the central density of the star
is equal to $\rho_c$ and the central pressure gradient vanishes.  Moreover, it
can be shown that the additional condition
\begin{equation}
	(m+1) (\psi_m(\xi_0)- \Psi_m(\xi_0))+\xi_0 (\psi'_m(\xi_0)- \Psi'_m(\xi_0))=0\label{add}
\end{equation}	
for $m\ge1$,
where $\xi_0$ is dimensionless radius of the unperturbed polytrope (i.e. 
$\Theta_0(\xi_0)=0$), should be set in order to have $\Theta$ vanishing on the 
perturbed stellar surface \citep{Ioka2001}. Eq.~(\ref{ausild0}) with the 
boundary conditions~(\ref{boundfint}) is simply the Lane-Emden equation 
for a polytrope of index $n$. Thus its solution $\Theta_0(\xi)$ is the density 
of the spherical, unmagnetized (i.e.\ unperturbed) star.

To summarize, the procedure to find the magnetically perturbed equilibrium of
the star is as follows:
Solve the unperturbed Lane-Emden Eq.~(\ref{ausild0}) for $\Theta_0$.
Choose the magnetic field's poloidal and toroidal components by specifying
$N_P$ and $N_T$ inside the star and boundary conditions relating to the field
just outside.
Then obtain the perturbed density profile by solving Eq.~(\ref{ausildm})
(see also \citep{Ioka2001,LIGOnote} for more details), subject to the boundary
conditions~(\ref{boundfint})--(\ref{add}).

\subsection{Perturbed quantities}

Here we give several useful integrals related to global properties of the
perturbed star.

In Newtonian gravity the addition of a magnetic field should not change the
mass of the star.
Therefore in general it changes the central density, for which we assume the
form
\begin{equation}
	\rho_c=\rho_0+\delta \rho_1 +{\cal O}(\delta^2).
	\label{rho}
\end{equation}
The first-order perturbed central density $\rho_1$ is found by writing the mass
\begin{eqnarray}
	M=C_M(M_0+\delta M_1 +{\cal O}(\delta^2))=\int_{V}\rho~d^3r,
	\label{massa}
\end{eqnarray}
where we remove dimensions using the constant
\begin{equation}
	C_M=4\pi \rho_0 \alpha^3_0.
	\label{CM}
\end{equation} 
Also, according to Eqs.~(\ref{newdefr}) and~(\ref{alpha}), we reference an
unperturbed characteristic length scale and radius of the star
\begin{equation}
	\alpha_0=R_0/\xi_0=\left[\frac{K(n+1)\rho_0^{-1+1/n}}{4\pi G}\right]^{1/2} .
	\label{alpha0}
\end{equation}
Using Eqs. (\ref{newdefrho}-\ref{newdefr}), (\ref{devtheta}) and (\ref{rho}) 
one has
\begin{eqnarray}
\nonumber	M=\int_{V}\rho_c\Theta^n(\xi)\alpha^{3}d^3\xi=4\pi \alpha^3_0\rho_0\int^{\xi_0}_{0}\xi^2\times\\\times(1+\delta \frac{\rho_1}{\rho_0} +{\cal O}(\delta^2))^{3/2n-1/2}(\Theta_{0}+\delta\Theta_{1}+{\cal O}(\delta^2))^n d\xi.
\end{eqnarray}
Thus, comparing with Eq.~(\ref{massa}), it can be shown that \citep{Ioka2001}
\begin{equation}
	M_0=\int^{\xi_0}_{0} \Theta^{n}_0(\xi) \xi^2 d\xi ,
	\label{mzero}
\end{equation}
while imposing the mass conservation condition $M_1=0$, yields 
\citep{Ioka2001}
\begin{equation}
	\frac{\rho_1}{\rho_0}=- \frac{2n^2}{M_0(3-n)}\int^{\xi_0}_{0} \xi^2 d\xi \psi_0(\xi)\Theta_0^{(n-1)}(\xi),
\end{equation}
where $\psi_0$ is defined in Eq.~(\ref{leg1}).


In view of the axisymmetry of the problem, we can write components of the
moment of inertia tensor (in units of $C_I=4\pi\rho_0\alpha^5_0$)
\begin{equation}
	{\cal I}_{11}={\cal I}_{22}=\frac{1}{2}({\cal I}_{11}+{\cal I}_{22})=\frac{1}{2C_I}\int_V \rho~(r^2+z^2) dx dy dz,
\end{equation}
where $(x,y,z)$ are the usual Cartesian coordinates.
Using Eqs. (\ref{newdefrho}-\ref{newdefr}) we have the dimensionless-coordinate
versions
\begin{equation}
	{\cal I}_{11}={\cal I}_{22}=\frac{1}{8\pi}\left(\frac{\rho_c}{\rho_0}\right)^{-\frac{3}{2}+\frac{5}{2n}}\int_V\Theta^n\xi^2(1+\cos^2\theta)d^3\xi,
\end{equation}
\begin{equation}
	{\cal I}_{33}=\frac{1}{4\pi}\left(\frac{\rho_c}{\rho_0}\right)^{-\frac{3}{2}+\frac{5}{2n}}\int_V\Theta^n\xi^2(1-\cos^2\theta)d^3\xi.
\end{equation}
Expanding up to first order in $\delta$,
\begin{equation}
	{\cal I}_{11}={\cal I}_{0}+\delta{\cal I}_{11,1}\label{expansI11}+{\cal O}(\delta^2),
\end{equation}
\begin{equation}
	{\cal I}_{33}={\cal I}_{0}+\delta{\cal I}_{33,1}\label{expansI33}+{\cal O}(\delta^2),
\end{equation}
it is possible to show that \citep{Ioka2001},
\begin{equation}
	{\cal I}_{0}=\frac{2}{3}\int^{\xi_0}_{0}\Theta^n_0\xi^4 d\xi,
	\label{Izero}
\end{equation}
\begin{equation}
	{\cal I}_{11,1}=\frac{2}{3}\left[\int^{\xi_0}_{0}n\Theta^{n-1}\left(\psi_0+\frac{1}{10}\psi_2\right)\xi^4d\xi\right]+\frac{5-3n}{2n}\frac{\rho_1}{\rho_0}{\cal I}_{0},	\label{i111}
\end{equation}
\begin{equation}
	{\cal I}_{33,1}=\frac{2}{3}\left[\int^{\xi_0}_{0}n\Theta^{n-1}_0\left(\psi_0-\frac{1}{5}\psi_2\right)\xi^4d\xi\right]+\frac{5-3n}{2n}\frac{\rho_1}{\rho}{\cal I}_{0},\label{i331}
\end{equation}
where $\psi_0$ and $\psi_2$ are defined according to Eq. (\ref{leg1}) and 
found by solving Eq. (\ref{ausildm}).


The total energy of a polytropic star with a magnetic field can be written as 
\citep[see e.g.\ ][]{Chandrasekhar1961,Woltjer1959}
\begin{equation}
{\cal E}={\cal M}+{\cal U}+{\cal W},
\label{enetot}
\end{equation}
where ${\cal M}$ is the magnetic energy, ${\cal U}$ is the internal energy and 
${\cal W}$ is the gravitational potential energy, that read 
\citep{Chandrasekhar1961}:
\begin{equation}
	{\cal M}=\frac{1}{8\pi C_E}\int_{V} |\vec{H}|^2 d^3r,
\end{equation}
\begin{equation}
	{\cal U}=\frac{n}{C_E}\int_{V}p d^3r,
\end{equation}	
\begin{equation}
	{\cal W}=-\frac{1}{2C_E}\int_{V}\rho \Phi d^3r,
\end{equation}
where we remove dimensions with the characteristic energy
\begin{equation}
	C_E=4\pi K(n+1)\rho^{(1+1/n)}_0\alpha^3_0.
	\label{CE}
\end{equation}
For polytropic configurations in hydromagnetic equilibrium, the virial theorem 
also holds \citep{Chandrasekhar1961}:
\begin{equation}
	{\cal M}+\frac{3}{n}{\cal U}+{\cal W}=0,
	\label{virial}
\end{equation}
and thus the total energy of the configuration can be written as
\begin{equation}
	{\cal E}=-\frac{3}{n}{\cal U}+{\cal U}=\frac{n-3}{n}{\cal U}.
\end{equation}
Expanding to first order in $\delta$
\begin{equation}
	{\cal M}=\delta {\cal M}_1 + {\cal O}(\delta^2),
\end{equation}
\begin{equation}
	{\cal U}={\cal U}_0+ \delta {\cal U}_1 + {\cal O}(\delta^2),
\end{equation}
\begin{equation}
	{\cal W}={\cal W}_0+ \delta {\cal W}_1 + {\cal O}(\delta^2),
\end{equation}
it is possible to show that \citep{Ioka2001}
\begin{equation}
	{\cal M}_1=\frac{1}{4\pi}\int_V\left(-\frac{\widetilde{\omega}^2}{2}P_0\Delta_5 P_0 + \frac{\widetilde{\omega}^2}{2} T^{2}_{0}\right)={\cal M}_{1,P}+{\cal M}_{1,T},
	\label{enemag}
\end{equation}
with ${\cal M}_{1,P}$ and ${\cal M}_{1,T}$ being the energy in the poloidal 
and toroidal field components---respectively,
\begin{equation}
	{\cal U}_0=\frac{n}{5-n}\xi^3_0\left(\frac{d\Theta_0}{d\xi}(\xi_0)\right)^2,
	\label{eq72Ioka}
\end{equation}
\begin{equation}
	{\cal U}_1=-\frac{n}{3-n}{\cal M}_1,\label{um1}
\end{equation}
\begin{equation}
	{\cal W}_0=-\frac{3}{n}{\cal U}_0,
	\label{w0}
\end{equation}
\begin{equation}
	{\cal W}_1=\frac{n}{3-n}{\cal M}_1.
	\label{w1}
\end{equation}
Then the total energy of the equilibrium configuration to first order
in $\delta$ reads
\begin{equation}
	{\cal E}=\frac{n-3}{n}{\cal U}_0+\delta{\cal M}_1=\frac{n-3}{n}{\cal U}_0+\delta({\cal M}_{1,P}+{\cal M}_{1,T}).
	\label{totene}
\end{equation}

The magnetic helicity $H = \int_V d^3r\, \vec{A} \cdot \vec{H}$ is also useful.
(Here $\vec{A}$ is the magnetic vector potential.)
For the field configuration used here, $\vec{A} \cdot \vec{H} \propto
\tilde\omega^2 P_0T_0$ \cite{Woltjer1959} and thus the helicity can be written
(in physical units)
\begin{equation}
H = \frac{8\pi}{3} \alpha_0C_E \delta \int_0^{\xi_0} d\xi\, \xi^4 P_0T_0.
\label{helicity}
\end{equation}


\subsection{Choice of Field Configuration}

Here we describe our special choice of magnetic field configuration and the
consequent properties of equilibria.

Following \citep{Ioka2001}, we choose all equilibria to have magnetic field
configurations such that:
\begin{eqnarray}
	N_p(\widetilde{\omega}^2 P_0)=-\widetilde{\omega}^2 P_0,~~~ & ~~~N_T(\widetilde{\omega}^2 P_0)=\lambda~\widetilde{\omega}^2 P_0,
	\label{choice1}
\end{eqnarray}
where $\lambda$ is a constant. With this choice, the solutions for $P_0 (\xi)$ 
and $T_0 (\xi)$ are functions of the radial coordinate only 
\citep{Trehan1972,Woltjer1960} and satisfy (see Equations~(\ref{delta5}) and 
(\ref{omegaT})):
\begin{eqnarray}
	\Delta_5 P_0 + \lambda^2 P_0=\Theta^n_0, ~~~ & ~~~T_0=\lambda P_0.
	\label{choice1a}
\end{eqnarray}
That is, the functions $P$ and $T$ specifying the poloidal and toroidal field 
components are proportional to each other, with their ratio being constant 
inside the star.
This configuration is the polytropic version of the simplest choice of
magnetic field (other than force-free) applicable to incompressible stars
\citep{Chandrasekhar1956,Prendergast1956,Woltjer1959,Woltjer1960}.
Although such a simple solution is unlikely to be perfectly realized in real
magnetars, its study has long been considered useful to give rough estimates of
the influence of the density gradient on the magnetic field.

\begin{figure*}
\includegraphics[width=8cm]{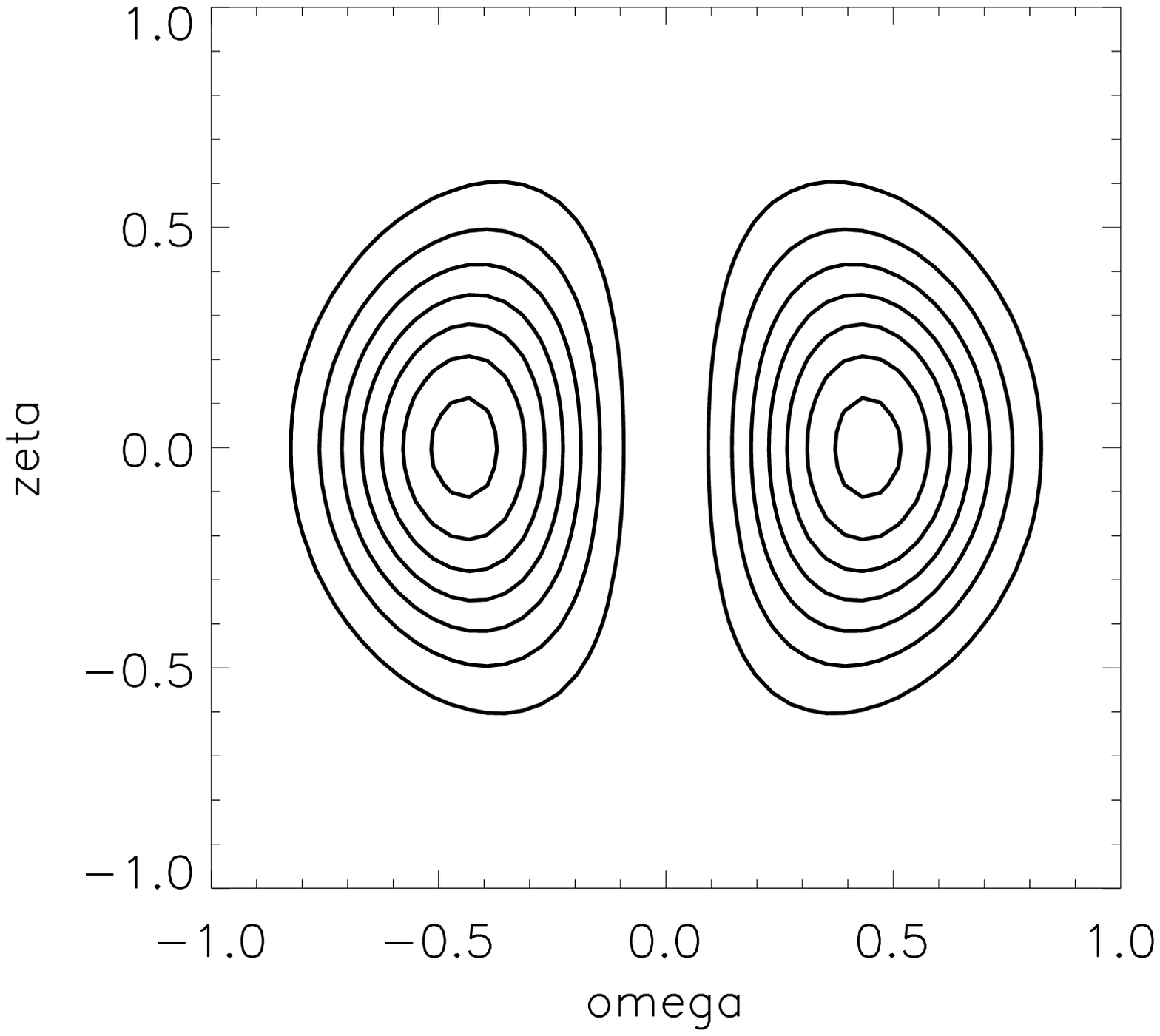}
\includegraphics[width=8cm]{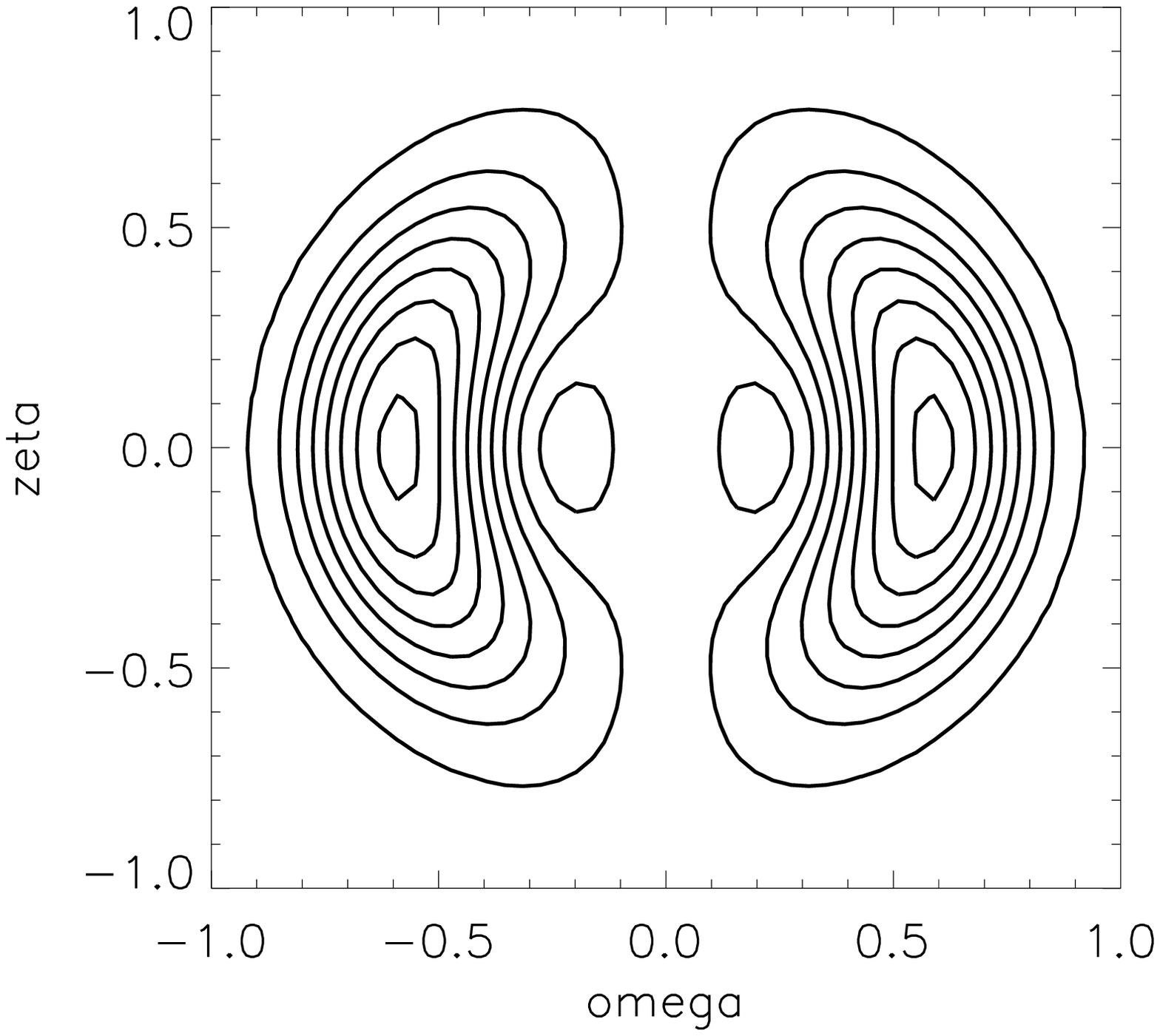}
\includegraphics[width=8cm]{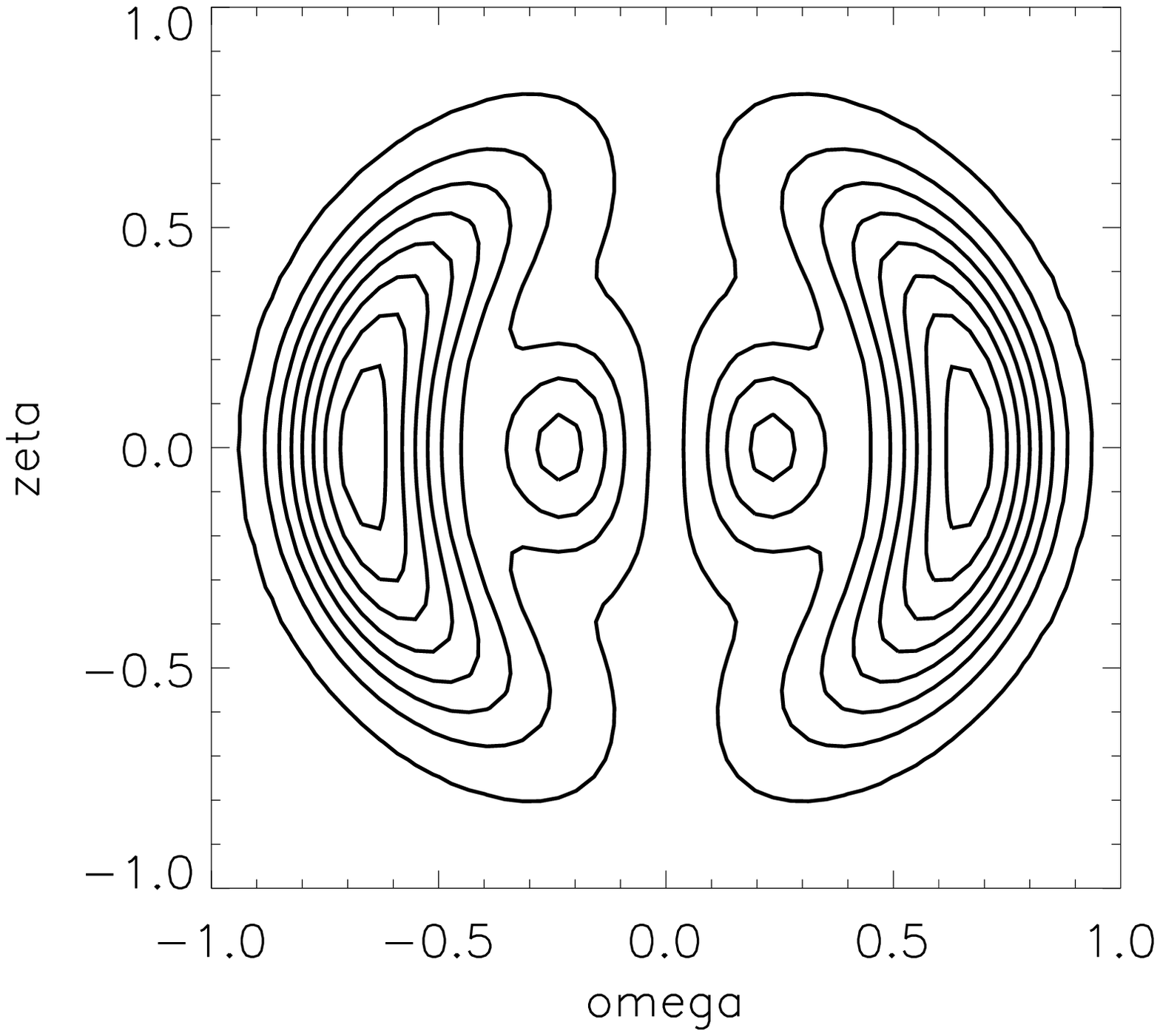}
\includegraphics[width=8cm]{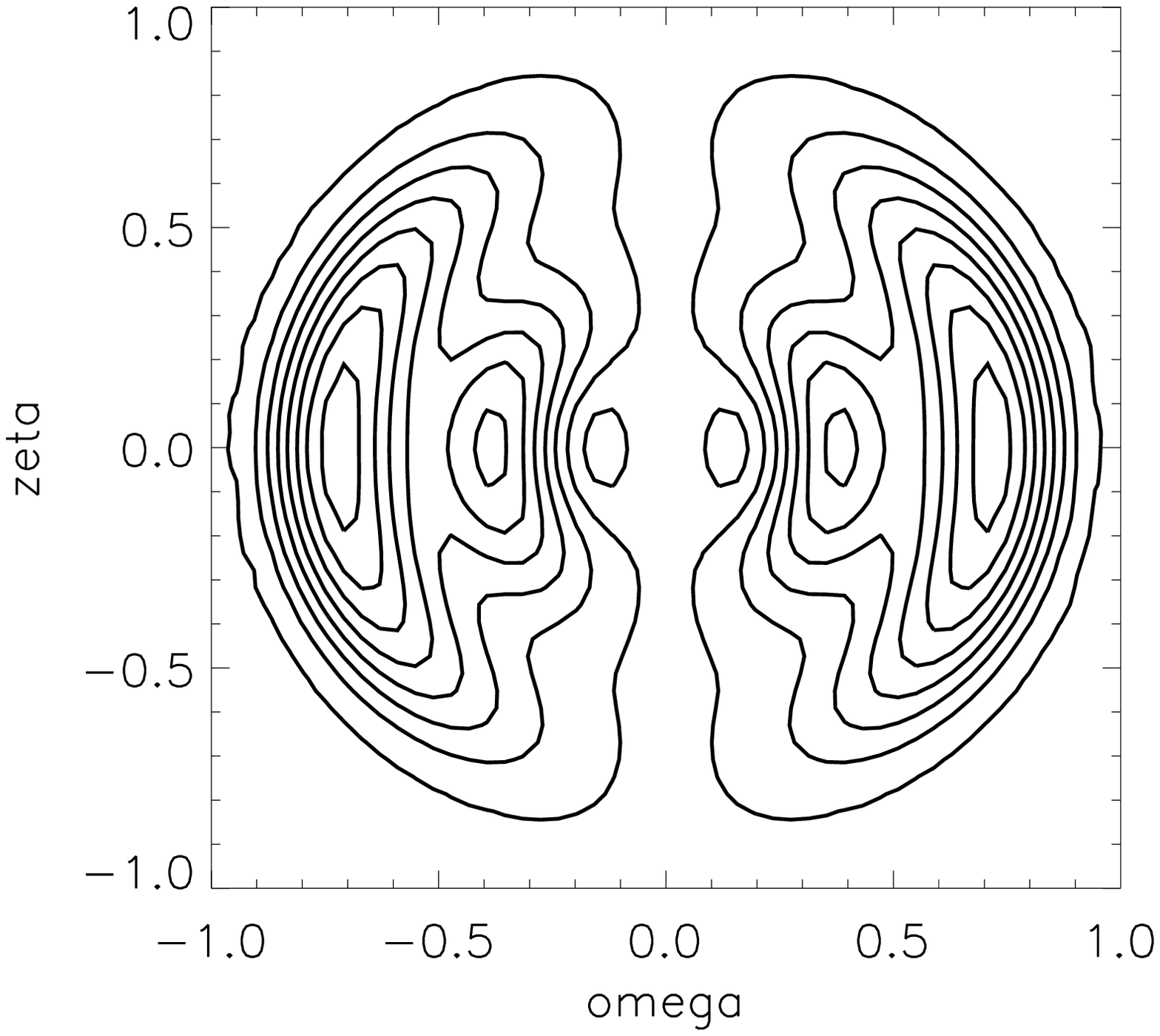}
\caption{Projection of the magnetic field lines on the meridional planes for 
the case of an $n=1$ polytrope and a magnetic field configuration 
characterized by an eigenvalue $\lambda_1$ (upper-left), $\lambda_2$ 
(upper-right), $\lambda_3$ (lower-left), $\lambda_4$ (lower right). 
\label{Fig2}}
\end{figure*}

Since the external magnetic field is expected to be 
negligible with respect to the internal one, boundary conditions are set so as 
to have the magnetic field vanishing on the star's surface (see also Section \ref{boundary}):
\begin{eqnarray}
	P_0(\xi_0)=0, ~~~ & ~~~ \frac{d P_0}{d\xi}(\xi_0)=0.
	\label{choice2}
\end{eqnarray}
Equation~(\ref{choice1a}) with the boundary conditions in Eq. (\ref{choice2})
gives \citep{Trehan1972,Woltjer1960}
\begin{eqnarray}
	\nonumber P_0(\xi)=\frac{\lambda}{\xi}n_1(\lambda\xi)\int^{\xi}_{0}\Theta^n_0(\xi')j_1(\lambda \xi') \xi'^3 d\xi'+\\+\frac{\lambda}{\xi}j_1(\lambda\xi)\int^{\xi_0}_{\xi}\Theta^n_0(\xi')n_1(\lambda \xi') \xi'^3 d\xi',
	\label{pzero}
\end{eqnarray}
where $j$ and $n$ are spherical Bessel and Neumann functions respectively, and
with $\lambda$ constrained to be a zero of the function:
\begin{equation}
	F(\lambda)=\int^{\xi_0}_{0}\Theta^n_0(\xi')j_1(\lambda \xi')\xi'^3 d\xi'.
	\label{lambdaroots}
\end{equation}
The first ten zeros of Eq.~(\ref{lambdaroots}) are indicated in the second 
column of Table\,\ref{Tab1}. These correspond 
to different magnetic field configurations, as shown in Fig. \ref{Fig2}, where 
we plot the magnetic field lines in the meridional planes (which run along the 
contours $\widetilde{\omega}^2 P_0(\xi)=const$), for the first four $\lambda$ 
roots of a $n=1$ polytrope. It is evident that the higher is $\lambda_k$, the 
more complex are the magnetic field lines. As commented in Section \ref{Sec:Justification}, 
also in light of the conditions required for the actual stability of the 
equilibrium state, in our analysis we consider configurations 
corresponding to the first ten $\lambda$ roots,
so as to deal with toroidal magnetic fields storing a ratio of the total magnetic energy
which is in between $\sim 65\%$ and $\sim 96\%$. 

\begin{table*}
	\begin{center}
		\begin{tabular}{cccccccc}
		k & $\lambda_k$ & ${\cal M}_1$ & ${\cal M}_{1,P}$& ${\cal M}_{1,T}/{\cal M}_{1,P}$& ${\cal I}_{11;1}/{\cal M}_1$ & ${\cal I}_{33;1}/{\cal M}_1$& ${\cal I}_{33;1}/{\cal M}_{1,P}$ \\
		\hline
		$1$ & 2.3619330 & 1.30707 & 0.45655454    &    1.86290&4.09418 & 2.27235 &        6.5055109   \\
		$2$ & 3.4078650   & 0.307662 & 0.078042433  &     2.94224&5.39981 & 0.881646 &       3.4756601  \\
		$3$ & 4.4300770   & 0.132171  & 0.024607442  &     4.37118&6.18915 & -0.314410&       -1.6887527   \\
		$4$ & 5.4434620   & 0.0734761 & 0.010269885  &     6.15452&6.70086 & -1.18614 &      -8.4862624  \\
		$5$ & 6.4524750   & 0.0470193 & 0.0050594294 &     8.29340& 7.04562 & -1.80598 &      -16.783694   \\
		$6$ & 7.4589800   & 0.0328329  & 0.0027852344 &     10.7882 &7.28585 & -2.25076 &      -26.532409  \\
		$7$ & 8.4639040   & 0.0243167 & 0.0016610675  &     13.6392&7.45842 & -2.57605 &     -37.711311  \\
		$8$ & 9.4677640   & 0.0187852  & 0.0010526103 &    16.8463 &7.58583 & -2.81905 &     -50.309612   \\
		$9$ & 10.470870   & 0.0149784  & 0.00069960812  &    20.4097  &7.68219 & -3.00433 &   -64.321804   \\
		$10$ & 11.473430  & 0.0122402  & 0.00048324273 &     24.3293 &7.75663 & -3.14830 &    -79.744235  \\
		\end{tabular}
\caption{For a $n=1$ polytrope, and different state indices $k$ (column 1) corresponding to 
the different eigenvalues $\lambda_k$ (column 2), we give: the first order (adimensional) magnetic energy (column 3), the 
(adimensional) poloidal field energy (column 4), the toroidal-to-poloidal 
magnetic energy ratio (column 5), the first order corrections to the moment of 
inertia tensor (per unit magnetic energy, columns 6-8). The values in columns 2, 3, 
5-7 are directly taken from \citet{Ioka2001}. 
\label{Tab1}}
\end{center}
\end{table*}

Using Eq.~(\ref{enemag}), we can compute for each equilibrium state 
characterized by a given $\lambda_k$ the corresponding values of the 
dimensionless total magnetic energy, the fraction of this energy going into the 
poloidal component, and the toroidal-to-poloidal energy ratio. These are 
listed in columns 3--5 of Table~\ref{Tab1}. As evident from such a Table, the 
higher is the value of $\lambda_k$, the higher is the fraction of energy 
stored in the toroidal field component.

Looking at columns 6-7 in Table \ref{Tab1}, it is evident that the corrections 
in the moment of inertia normalized to the total magnetic energy of the state, 
are such that the higher is $\lambda_k$, the more prolate is the star. This is 
equivalent to say that states having the same total magnetic energy but higher 
toroidal-to-poloidal field energy ratio, are more prolate. Physically, this is 
a consequence of the fact that the toroidal field tends to make the star 
prolate, working like a rubber belt tightening up the equator of the star.

\section{Generalization and results}
\label{Sec:Results}
\citet{Ioka2001} has invoked jumps between the different equilibrium configurations
of a magnetized neutron star to explain the properties of SGR flares.
Here we explore the model \citep{Ioka2001} in terms of flare
observables: jumps in energy and moment of inertia.
First, we describe the choice of jumps considered by \citet{Ioka2001}
(conserving the total magnetic energy and requiring $\Delta {\cal I} /{\cal I}=10^{-4}$). Next, we present results for a new choice of jumps (conserving the energy of the poloidal field only). Finally, we discuss the dependence on the mean 
poloidal field strength for jumps that conserve the poloidal field energy,
and describe the uncertainties associated with using a set of stellar
models with $n=1$ polytropic EOS.

\subsection{Jump conditions}

Equilibria of non-magnetic polytropes can be characterized by one parameter,
e.g.\ the gravitational potential energy; while equilibria of magnetic
polytropes require two, e.g.\ the gravitational and magnetic potential
energies. These two degree of freedom also allow one to choose the the two
observables of SGR flares, total energy and moment of inertia, as parameters of the problem.
Considering jumps between equilibria of a single star requires fixing the
mass, leading to a sequence of equilibria characterized by a single parameter,
e.g.\ the ratio of potential energies $\delta$.
Therefore jumps between equilibria, which are to model SGR flares, can trace
various paths in the two-dimensional parameter space.

\begin{figure*}
\includegraphics[width=9cm,angle=90]{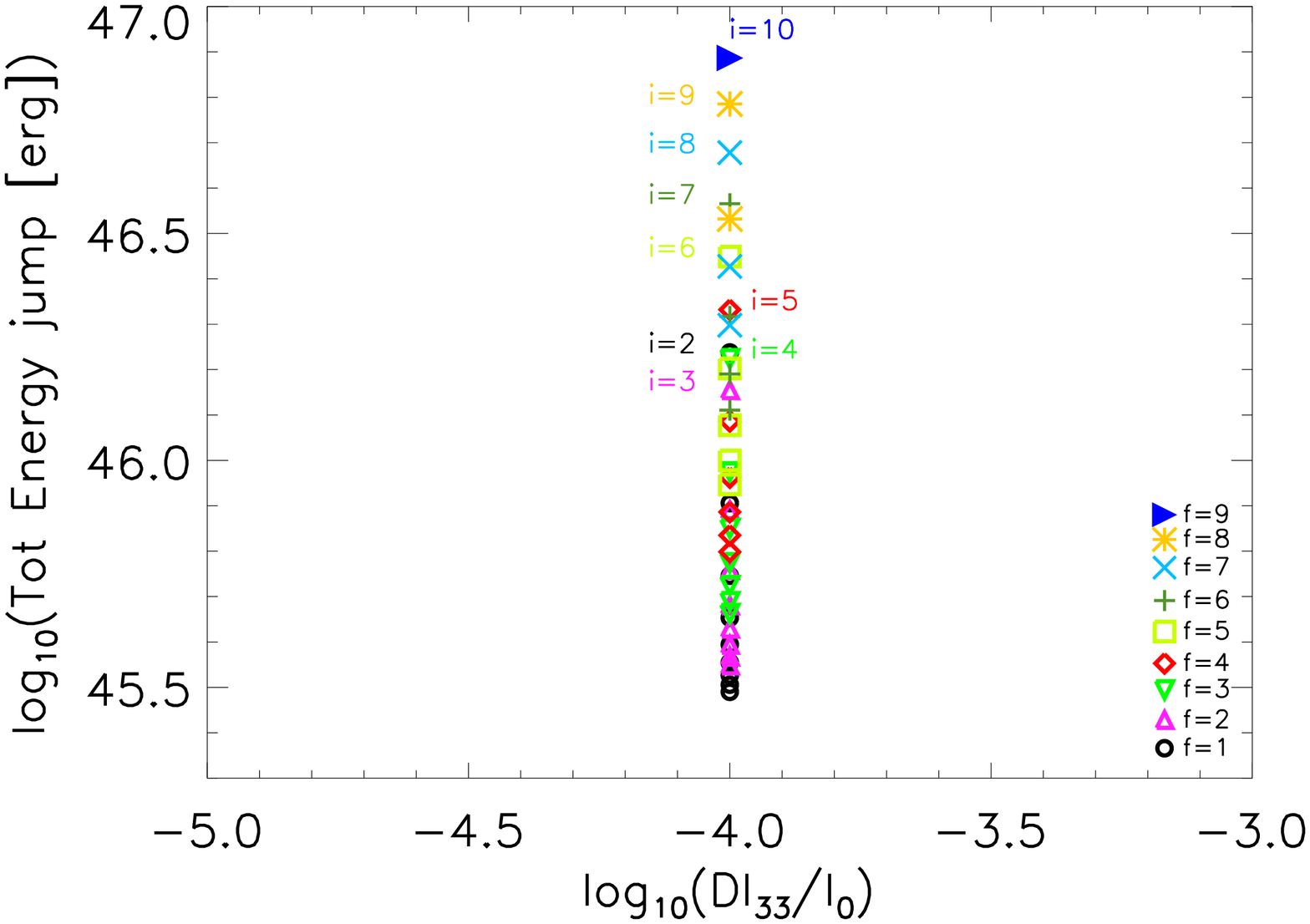}
\caption{Energy vs moment of inertia jumps for different final ($f$)
and initial ($i$) state indices in jumps conserving the total
magnetic energy (to first order), and with a change in moment of inertia of $10^{-4}$ 
 (as possibly observed in the August giant flare of SGR\,1900+14, see \citep{Ioka2001}). Jumps characterized by
the same $f$ are plotted with the same color and symbol. For clarity, for each $f$, we mark on the plot the initial
state index i of the jump with the highest energy, corresponding to $i=f+1$. \label{jumps_ioka}}
\end{figure*}

In Fig.~\ref{jumps_ioka}, we plot the paths traced by the specific families of
jumps considered by \citet{Ioka2001}. For this family, $\Delta {\cal I}/{\cal I}=10^{-4}$ and the total magnetic energy is kept constant in a jump. Because of this last requirement, since the contribution from the toroidal field decreases in a jump (see column 5 in Table\,\ref{Tab1}), the poloidal field increases. Because the toroidal fields make the star more prolate, and poloidal fields do the reverse, this allows a large change in the moment of inertia. 

Note that in Ioka's model, for a given value of the final state index $f$,
jumps from initial states with higher values of initial state index $i$ release
a smaller amount of energy. This is due to the fact that, for increasing $i$,
the ratio of toroidal-to-poloidal field energy increases. Thus, higher values
of $i$ require a lower value of the total magnetic energy in the star if a
fixed moment of inertia change is required in all jumps $i$-to-$f$ with the
same $f$.  This in turn implies a smaller jump in total energy with higher
values of $i$, the energy jump being proportional to the square of the
magnetic-to-gravitational potential energy ratio of the initial state (see
Eq.~(94) of \citet{Ioka2001}).

In the present work we modify the calculations by \citet{Ioka2001} by proposing
a second family of higher-energy jumps based on keeping the potential energy of
the poloidal magnetic field constant.
The calculation of \citet{Ioka2001} is mainly modified in the fact that, since
we allow the magnetic energy to change in jumps, we only need first order
perturbation theory, while \citet{Ioka2001} needed second order. 
Besides the fact that larger energy jumps are obtained allowing the total
magnetic energy to change, our choice is physically interesting for two
reasons.
First, in real magnetars, the internal poloidal field may remain matched to the
outer poloidal field, which does not change by of order unity even in giant flares.
Second, our choice is consistent with the standard theory that magnetic
helicity is expelled from the star \cite{Thompson2002}, since the helicity
decreases through jumps (see below).
This is more desirable than the behavior of Ioka's model, where the helicity
increases in lower energy states.
 
Consider a transition $(i,f)$ between two equilibrium states, the initial being
characterized by an eigenvalue $\lambda_i$, the final by $\lambda_f$.
This means the magnetic to gravitational potential energy will, unlike the
case of \citet{Ioka2001}, have different initial and final values $\delta_i$
and $\delta_f$.
If we make the hypothesis that the energy in the poloidal field is conserved in
the transition, then the following relation holds:
\begin{equation}
	\delta_i{\cal M}_{1,P}(\lambda_i)=\delta_f{\cal M}_{1,P}(\lambda_f).
	\label{conspol}
\end{equation}
As evident from Table\,\ref{Tab1}, ${\cal M}_{1,P}(\lambda_i)<{\cal M}_{1,P}(\lambda_f)$ for $i>f$. Thus, while in Eq.~(\ref{helicity}) the integral increases at lower-numbered states, the choice in Eq.\,(\ref{conspol}) assures $\delta_f<\delta_i$ for $i>f$, making the overall helicity decrease in this family of jumps.

Using Eq.~(\ref{totene}) and the above condition, the total energy change in 
the transition reads
\begin{eqnarray}
	\nonumber \Delta{\cal E}_{(i,f)}=\delta_f{\cal M}_{1,T}(\lambda_f)-\delta_i{\cal M}_{1,T}(\lambda_i)=\\=\delta_i{\cal M}_{1,P}(\lambda_i)\left[\frac{{\cal M}_{1,T}(\lambda_f)}{{\cal M}_{1,P}(\lambda_f)}-\frac{{\cal M}_{1,T}(\lambda_i)}{{\cal M}_{1,P}(\lambda_i)}\right].
\end{eqnarray}
Looking at the 5th column in Table \ref{Tab1}, it is evident that to power an 
SGR flare (i.e.\ $\Delta {\cal E}_{(i,f)}< 0$), only jumps from higher to lower 
$\lambda_k$ are permitted. As we will show in the next section, physically 
this corresponds to having the star becoming less prolate (i.e.\ more 
spherical) in the transition, thus passing from a more energetic to a less 
energetic equilibrium configuration.

To completely specify the energy (in physical units) of equilibria, three 
parameters are needed: two of them pertain the EOS (e.g.\ the 
total mass $M=C_M M_0$ and the unperturbed radius $R_0=\alpha_0\xi_0$), while 
the third is the ratio $\delta$ between the physical unit in which we measure 
the gravitational potential energy (that is fixed by $M$ and $R_0$) and the 
magnetic energy. For a star characterized by a given $M$ and $R_0$, a 
transition $(i,f)$ leaves us with two parameters: the values of 
$\delta_i$ and $\delta_f$. The requirement of having the poloidal field energy 
conserved in the jump fixes $\delta_f$ as a function of $\delta_i$ (see Eq. 
(\ref{conspol})) and leaves only $\delta_i$ free. Rather than specifying the 
last, we can equivalently specify the strength of the mean poloidal magnetic 
field inside the star,
\begin{equation}
	\left<H_{P}\right>=\sqrt{8\pi \frac{C_{E}\delta_i {\cal
M}_{1,P}(\lambda_i)}{(4\pi R^3_0/3)}},
\end{equation}
and thus the energy jumps (in physical units) are given by
\begin{equation}
	C_E \Delta{\cal E}_{(i,f)}=\frac{\left<H_{P}\right>^2}{8\pi}\frac{4\pi R^3_0}{3}\left[\frac{{\cal M}_{1,T}(\lambda_f)}{{\cal M}_{1,P}(\lambda_f)}-\frac{{\cal M}_{1,T}(\lambda_i)}{{\cal M}_{1,P}(\lambda_i)}\right]
	\label{dene}
\end{equation}

In a transition $(i,f)$ between two equilibrium states that conserves the 
poloidal field energy inside the star, the moment of inertia changes as
\begin{eqnarray}
	\nonumber \frac{\Delta {\cal I}_{33,(i,f)}}{{\cal I}_0}=\delta_f \frac{{\cal I}_{33,1}(\lambda_f)}{{\cal I}_0}-\delta_i \frac{{\cal I}_{33,1}(\lambda_i)}{{\cal I}_0}=\\=\frac{\left<H_{P}\right>^2}{8\pi}\frac{4\pi R^3_0}{3 C_E {\cal I}_0 }\left[\frac{{\cal I}_{33,1}(\lambda_f)}{{\cal M}_{1,P}(\lambda_f)}-\frac{{\cal I}_{33,1}(\lambda_i)}{{\cal M}_{1,P}(\lambda_i)}\right].
\label{jumpinerz}
\end{eqnarray}

\begin{figure*}
\includegraphics[width=9cm,angle=90]{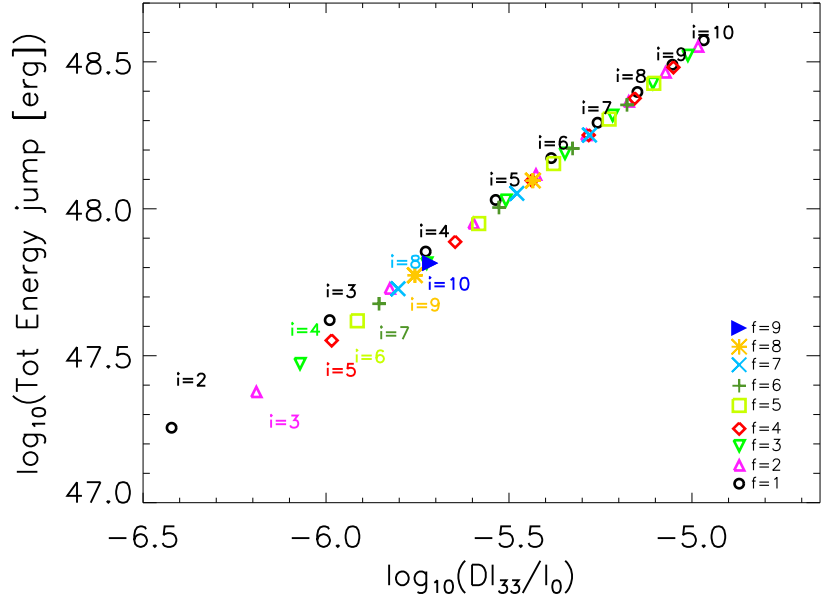}
\caption{
Energy vs moment of inertia jumps for different final ($f$)
and initial ($i$) state indices, for a family of jumps that conserves the poloidal field
strength ($10^{15}$~G). Colors are as in Fig. \ref{jumps_ioka}. We mark on
the plot the initial state index of all the jumps with $f=1$.
For jumps with $f>1$, we mark for clarity only the jump with
lowest energy, corresponding to $i=f+1$.
\label{jumps_ours}}
\end{figure*}

In Fig.~\ref{jumps_ours}, we plot our family of fixed poloidal magnetic
energy jumps. As evident from this Figure, keeping the poloidal magnetic field
constant, the resulting energy jumps range in between $\sim 2\times10^{47}$
and $\sim 4\times10^{48}$ erg, while moment of inertia jumps are always
$\lesssim 10^{-4}$ (the upper limit observed in the 1998 giant flare of
SGR\,1900+14). 
Higher energy jumps are possible, but require higher jumps in the moment of
inertia. We note, however, that smaller changes in $\Delta {\cal  I}/{\cal I}$ (i.e. in the observed spin period) 
could also be produced by a magnetic field axis misaligned with the rotation axis.
Note also that, for the family of jumps we have considered here, the 
total internal magnetic field is of the order of 1--2$\times10^{16}$~G, smaller 
than required by \citet{Ioka2001}. In his model, in fact, the
total internal magnetic field is $\gtrsim 10^{17}$\,G for jumps with energies
$\gtrsim 10^{48}$\,erg (see Figs.~3 and 4 in \citep{Ioka2001}).

The fundamental result here is that our choice of jumps is particularly large in energy and small in moment of inertia. 
In fact, allowing for a change in total magnetic energy, produce energies larger than those of \citet{Ioka2001} (see Fig.~\ref{jumps_ioka}) by $O(1/\delta)$. Moreover, since $\lambda > 1$ always, the toroidal field energy dominates in the equilibria. Because our family of jumps only conserves the poloidal field energy, they can change by a significant fraction the total magnetic energy. On the other hand, our moment of inertia changes are smaller than for Ioka's choice (see Fig.~\ref{jumps_ioka}) since, as noted above, the decrease in toroidal field and increase in poloidal field in Ioka's model tend to add up their effect in increasing the moment of inertia (making the star less prolate).
 
\subsection{Poloidal field energy dependence}
\begin{figure}
\begin{center}
\includegraphics[width=8.5cm]{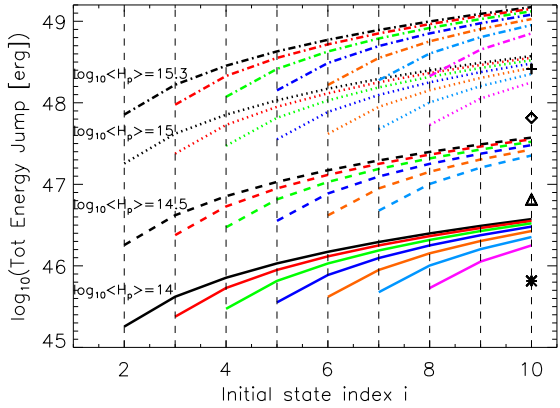}
\caption{Total energy jumps as a function of the 
initial state index $i$ for final states having indices $f=1$ (black lines), 
$f=2$ (red lines), $f=3$ (green lines), $f=4$ (blue lines), $f=5$ (orange 
lines), $f=6$ (light blue lines), $f=7$ (purple lines), $f=8$ (yellow lines), 
$f=9$ (black symbols). The jumps are computed for different values of the mean 
poloidal field $\left<H_{P}\right>$, which is conserved in the transition: from 
bottom to top, $10^{14}$ G (solid lines and asterisk), $10^{14.5}$ G (dashed 
lines and triangle), $10^{15}$ G (dotted lines and diamond), $10^{15.3}$ G 
(dot-dashed lines and cross). A $n=1$ polytrope with $M=C_M M_0=1.4 M_{\odot}$ 
and $R_0=10^6$ cm is being considered. \label{Fig3}}
\end{center}
\end{figure}

To show the effect of the poloidal magnetic field strength on the family jumps introduced in the previous Section,
in Fig.~\ref{Fig3} we show, for a $n=1$ polytrope with $R_0=\alpha_0 
\xi_0=10^6$ cm, the energy jumps $-C_E \Delta{\cal E}_{(i,f)}$ as a function 
of the index $i$ of the initial state, for final states $f=1-9$, and 
$\left<H_{P}\right>=(10^{14}, 10^{14.5},10^{15},10^{15.3})$ G. These values of 
the mean poloidal field correspond to a total mean magnetic field inside the 
star lower than $\approx 10^{16}$ G, for transitions 
having $f<10$ (see column 5 in Table \ref{Tab1}).

\begin{figure}
\begin{center}
\includegraphics[width=8.5cm]{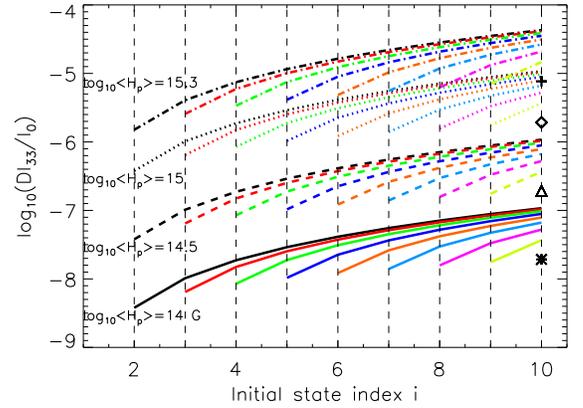}
\caption{Moment of inertia jumps, $\Delta{\cal I}_{33}/{\cal I}_0$, as a 
function of the initial state index $i$ for different final states and 
different values of the mean poloidal field strength (see caption of Fig. 
\ref{Fig3} for colors and symbols). A $n=1$ polytrope with $M=1.4 M_{\odot}$ 
and $R_0=10^6$ cm is being considered. \label{Fig4}}
\end{center}
\end{figure}

In Fig. \ref{Fig4} we show, for a $n=1$ polytrope with $M_0=1.4 M_{\odot}$ and 
$R_0=10^6$~cm, the moment of inertia jumps $\Delta {\cal I}_{33,(i,f)}/{\cal 
I}_0$ as a function of the index $i$ of the initial state, for final states 
$f=1-9$, and $\left<H_{P}\right>=(10^{14}, 10^{14.5},10^{15},10^{15.3})$~G. We 
note that in all cases $\Delta {\cal I}_{33,(i,f)}/{\cal I}_0<10^{-4}$, i.e.\
the transitions considered here are all associated with changes in moment of
inertia smaller than the possible value inferred from the 1998 August 17
giant flare from SGR\,1900+14 by \citet{Ioka2001}.
Small jumps in moment of inertia can be hidden by the high timing noise and
sparse observations of magnetar spin periods:
For example, a jump of $5\times10^{-6}$ could have happened in the 2004 giant
flare of SGR~1806$-$20 \citep{WoodsEtAl2007}.

\subsection{Equation of State Dependence}

The EOS is the simplifying assumption which seems quantitatively
most important in Ioka's calculations.
Fig.~3 of \citet{Ioka2001} shows that the highest jumps in energy are found
for $n=2.5$ polytropes, extremely soft EOS on the verge of
being unstable to radial perturbations; and that energies for the more
realistic $n=1$ polytropes are orders of magnitude lower.

In contrast to \citet{Ioka2001}, we schematically examine the EOS-dependence of
GW energy by restricting the polytropic index to $n=1$ and varying the mass and
radius of the star instead.
For many problems this approach gives numbers which are comparable to those for
more realistic EOS.
Increasing the polytropic index as in \citet{Ioka2001} can lead to artificially
large energy jumps, as the star approaches instability to radial perturbations
as $n\to3$.
(e.g. \citep{Chandrasekhar1967, Ledoux1946, Chan1968}). 
Also, as evident from Fig.~\ref{Fig2}, the magnetic field is usually concentrated in the
outer core of the star (densities at or slightly above nuclear density), where
all realistic EOS tend to be fit well by polytropes with
$n=0.5$--1 \citep{Read2008}.
As \citet{Ioka2001} showed, the energy jumps tend to rise with $n$, so $n=1$ is
good for estimating the maximum energy.

The choice $n=1$ also makes the math very simple: we have
\citep{Chandrasekhar1961}
\begin{eqnarray}
	\Theta_0(\xi)=\frac{\sin\xi}{\xi}, ~~~ & ~~~ \xi_0=\pi. ~~~ &
	\label{plochoice}
\end{eqnarray}	
Dimensionless unperturbed quantities are then easily calculated:
\begin{equation}
	M_0=\int^{\pi}_{0}(\sin\xi/\xi)\xi^2d\xi=\pi\,,
\end{equation}
from Eq.~(\ref{mzero});
\begin{equation}
	{\cal I}_0=\frac{2}{3}\int^{\pi}_{0}\xi^3 \sin\xi
	d\xi=\frac{2}{3}(\pi^3-6\pi)\,,
\end{equation}
from Eq.~(\ref{Izero}).
For a choice of (dimensionful) mass $M=C_M M_0$ and radius $R_0=\alpha_0 \xi_0$
of the unperturbed star, the dimensionful conversion factors are derived as
\begin{equation}
	C_M=M/\pi\,,
\end{equation}
from Eq.~(\ref{massa}); 
\begin{equation}
	\alpha_0=R_0/\pi\,, 
\end{equation}
from Eq.~(\ref{alpha0});
\begin{equation}
	K=2\pi G\alpha^2_0\,,
\end{equation}
from Eq.~(\ref{alpha0}); 
\begin{equation}
	\rho_0=C_M/(4\pi\alpha^3_0)\,,
\end{equation}
from Eq.~(\ref{CM}).
Inserting in Eq.~(\ref{CE}) we derive the energy scale for our results in
physical units,
\begin{equation}
C_E=\frac{G C^2_M}{\alpha_0}=\frac{G M^2}{\pi R_0}
	\label{cescal}
\end{equation}
which, for the canonical choice $M=1.4 M_{\odot}$ and $R_0=10^6$~cm, yields
$C_E\sim 1.6\times10^{53}$~erg.

To estimate the mass and radius dependence of the results we pick ranges of
the parameters based on observations.
The present observed mass range is roughly 1.2--2.0~$M_\odot$ (see \citep{Demorest:2010bx} for the highest mass)
and predicted radii are roughly 9--15~km (for a summary see \citep{Read2008}).

The energy jumps scale differently between Ioka's jump condition and the
constant poloidal field condition.
In this last case the magnetic energy, which for a given equilibrium state scales as $\left<H_P\right>\times R^{3}_{0}$,
 is the source of the flare (i.e.\ jump, see Eq.~(\ref{dene})). Thus, for a fixed value of
 $\left<H_P\right>$, the larger is the star's radius, the higher is the energy jump.
For $R_0=15$~km, the energy jumps shown in Fig.~\ref{Fig3} would be a factor of
$\approx 3.4$ higher than for $R_0=10$~km.

On the other hand, the transitions considered by \citet{Ioka2001} 
involved no change in the total magnetic energy.
In this case, the energy source is the gravitational potential energy, so the
total energy jumps scaled as $C_E\propto M^2/ R_0$, with higher masses and
smaller radii favoring more powerful flares.
For $M=2 M_{\odot}$ and $R_0=9$ km, this scaling increases the energy jumps
by a factor of $\approx 2.3$ with respect to a standard choice of
$M=1.4~M_{\odot}$ and $R_0=10$~km.

Concerning the (fractional) moment of inertia jumps, in \citet{Ioka2001} they were fixed to match the
value derived for the August giant flare of SGR\,1900+14. But
using the fixed-poloidal jump condition they can change.
We see from Eqs.~(\ref{jumpinerz}) and~(\ref{cescal}) that, for a given
$\left<H_P\right>$, we have $\Delta {\cal
I}_{33,(i,f)}/{\cal I}_0 \propto R^{2}_0/M^{3}_0$. Thus, bigger masses and smaller radii 
help to keep the moment of inertia jumps small in a transition.
In particular, for $M=2~M_{\odot}$ and $R_0=9$~km, the jumps shown in
Fig.~\ref{Fig4} are reduced by a factor of $\approx 3.6$.


\section{Discussion}
\label{Sec:Discussion}

We have shown that changes in the hydromagnetic deformation of a magnetar can
provide an energy reservoir of order $10^{48}$--$10^{49}$~erg, comparable to
LIGO and Virgo observational upper limits on $f$-mode GW emission, under more
generic circumstances than considered in the original work by
\citet{Ioka2001}.
The key requirement is a change in the magnetic potential energy of the star,
which causes the change in total energy of the star to be first-order rather
than second-order in the hydromagnetic perturbation parameter.
Such an event can happen of order ten times over the lifetime of the star, and
such energies are then only applicable to (some of) the rare giant flares.
However, in the family of jumps we proposed here to explain SGR flares, 
a large glitch in the magnetar's spin is not required, nor is an
unrealistically soft EOS or extremely high internal field.
Our family of jumps is also consistent with the idea that the helicity of the
internal field is decreased rather than increased in giant flares.

We have briefly noted that such high energies are also available in the
standard model of magnetar flares, crust cracking, if the solid part of the
star is not limited to the crust but includes a core of solid quarks or
mixed-phase material.

We have only considered equilibrium states and the total energy available.
Our estimates are order of magnitude accuracy, and could be carried further by
considering refinements such as relativity, field configurations, and
realistic EOS.
To establish high GW emission energies as a viable model also requires
investigation of the dynamics to determine if the ratio of GW/EM energy
emitted can be much higher than unity, for example if most of the action takes
place in the interior of the star.

We conclude by noting that the problem of GW emission from magnetar flares
presents further opportunities:
It is a problem that has received much less study than, for example,
continuous GW emission from newborn magnetars.
Yet many of the those results can be adapted to this problem.
And, while newborn magnetars may become relevant to observations in the era of
advanced interferometers, the flare problem is relevant right now.
We hope that this will spur further work on the problem.

\section*{Acknowledgments}

We are grateful to R. Rutledge and to many members of the LIGO Scientific
Collaboration for helpful discussions, especially D.I. Jones, P. Kalmus, Yu.\ 
Levin, S. Marka, M. Papa, and D. Reitze.
This work was partially supported by the ``Fondazione Angelo della Riccia'' - bando A.A.
2008--2009 (AC); by NSF grants PHY-0555628 and PHY-0855589, and by the LIGO Visitors Program (BJO).
LIGO was constructed by the California Institute of Technology and
Massachusetts Institute of Technology with funding from the National Science
Foundation and operates under cooperative agreement PHY-0757058.
This paper has document number LIGO-P1100011.
AC gratefully acknowledges the Penn State Institute for Gravitation and the
Cosmos and the Albert Einstein Institute in Hannover for partially
supporting this project during its initial and middle stages, respectively, 
and thanks F. Ricci for encouragement.

\bibliography{ACBO09bib}

\begin{thebibliography}{95}
\expandafter\ifx\csname natexlab\endcsname\relax\def\natexlab#1{#1}\fi
\expandafter\ifx\csname bibnamefont\endcsname\relax
  \def\bibnamefont#1{#1}\fi
\expandafter\ifx\csname bibfnamefont\endcsname\relax
  \def\bibfnamefont#1{#1}\fi
\expandafter\ifx\csname citenamefont\endcsname\relax
  \def\citenamefont#1{#1}\fi
\expandafter\ifx\csname url\endcsname\relax
  \def\url#1{\texttt{#1}}\fi
\expandafter\ifx\csname urlprefix\endcsname\relax\def\urlprefix{URL }\fi
\providecommand{\bibinfo}[2]{#2}
\providecommand{\eprint}[2][]{\url{#2}}

\bibitem[{\citenamefont{Abbott et~al.}(2007)}]{Abbott:2007zzb}
\bibinfo{author}{\bibfnamefont{B.}~\bibnamefont{Abbott}} \bibnamefont{et~al.}
  (\bibinfo{collaboration}{LIGO Scientific Collaboration}),
  \bibinfo{journal}{Phys. Rev. D} \textbf{\bibinfo{volume}{76}},
  \bibinfo{pages}{062003} (\bibinfo{year}{2007}).

\bibitem[{\citenamefont{Abbott et~al.}(2008)}]{Abbott:2008gj}
\bibinfo{author}{\bibfnamefont{B.}~\bibnamefont{Abbott}} \bibnamefont{et~al.}
  (\bibinfo{collaboration}{LIGO Scientific Collaboration}),
  \bibinfo{journal}{Phys. Rev. Lett.} \textbf{\bibinfo{volume}{101}},
  \bibinfo{pages}{211102} (\bibinfo{year}{2008}).

\bibitem[{\citenamefont{Abbott et~al.}(2009)}]{Abbott:2009zd}
\bibinfo{author}{\bibfnamefont{B.~P.} \bibnamefont{Abbott}}
  \bibnamefont{et~al.} (\bibinfo{collaboration}{LIGO Scientific
  Collaboration}), \bibinfo{journal}{Astrophys. J.}
  \textbf{\bibinfo{volume}{701}}, \bibinfo{pages}{L68} (\bibinfo{year}{2009}).

\bibitem[{\citenamefont{Abadie et~al.}(2010)}]{Abadie:2010wx}
\bibinfo{author}{\bibfnamefont{J.}~\bibnamefont{Abadie}} \bibnamefont{et~al.}
  (\bibinfo{year}{2010}), \eprint{arXiv:1011.4079}.

\bibitem[{\citenamefont{Harry}(2010)}]{aLIGO}
\bibinfo{author}{\bibfnamefont{G.~M.} \bibnamefont{Harry}}
  (\bibinfo{collaboration}{for the LIGO Scientific Collaboration}),
  \bibinfo{journal}{Class. Quant. Grav.} \textbf{\bibinfo{volume}{27}},
  \bibinfo{pages}{084006} (\bibinfo{year}{2010}).

\bibitem[{\citenamefont{Acernese et~al.}(2009)}]{aVirgo}
\bibinfo{author}{\bibfnamefont{F.}~\bibnamefont{Acernese}} \bibnamefont{et~al.}
  (\bibinfo{collaboration}{Virgo Collaboration}) (\bibinfo{year}{2009}),
  \bibinfo{note}{{Virgo internal note VIR-0027A-09}},
  \urlprefix\url{https://tds.ego-gw.it/itf/tds/}.

\bibitem[{\citenamefont{{Kalmus} et~al.}(2009)\citenamefont{{Kalmus}, {Cannon},
  {M{\'a}rka}, and {Owen}}}]{Kalmus2009}
\bibinfo{author}{\bibfnamefont{P.}~\bibnamefont{{Kalmus}}},
  \bibinfo{author}{\bibfnamefont{K.~C.} \bibnamefont{{Cannon}}},
  \bibinfo{author}{\bibfnamefont{S.}~\bibnamefont{{M{\'a}rka}}},
  \bibnamefont{and} \bibinfo{author}{\bibfnamefont{B.~J.}
  \bibnamefont{{Owen}}}, \bibinfo{journal}{{Phys. Rev. D}}
  \textbf{\bibinfo{volume}{80}}, \bibinfo{pages}{042001}
  (\bibinfo{year}{2009}).

\bibitem[{\citenamefont{Lander and Jones}(2010)}]{LanderJones2011}
\bibinfo{author}{\bibfnamefont{S.~K.} \bibnamefont{Lander}} \bibnamefont{and}
  \bibinfo{author}{\bibfnamefont{D.~I.} \bibnamefont{Jones}}
  (\bibinfo{year}{2010}), \eprint{arXiv:1009.2453}.

\bibitem[{\citenamefont{{Mazets} et~al.}(1999)\citenamefont{{Mazets},
  {Aptekar}, {Butterworth}, {Cline}, {Frederiks}, {Golenetskii}, {Hurley}, and
  {Il'Inskii}}}]{Mazets1999}
\bibinfo{author}{\bibfnamefont{E.~P.} \bibnamefont{{Mazets}}},
  \bibinfo{author}{\bibfnamefont{R.~L.} \bibnamefont{{Aptekar}}},
  \bibinfo{author}{\bibfnamefont{P.~S.} \bibnamefont{{Butterworth}}},
  \bibinfo{author}{\bibfnamefont{T.~L.} \bibnamefont{{Cline}}},
  \bibinfo{author}{\bibfnamefont{D.~D.} \bibnamefont{{Frederiks}}},
  \bibinfo{author}{\bibfnamefont{S.~V.} \bibnamefont{{Golenetskii}}},
  \bibinfo{author}{\bibfnamefont{K.}~\bibnamefont{{Hurley}}}, \bibnamefont{and}
  \bibinfo{author}{\bibfnamefont{V.~N.} \bibnamefont{{Il'Inskii}}},
  \bibinfo{journal}{\apj} \textbf{\bibinfo{volume}{519}}, \bibinfo{pages}{L151}
  (\bibinfo{year}{1999}).

\bibitem[{\citenamefont{{Duncan} and {Thompson}}(1992)}]{Duncan1992}
\bibinfo{author}{\bibfnamefont{R.~C.} \bibnamefont{{Duncan}}} \bibnamefont{and}
  \bibinfo{author}{\bibfnamefont{C.}~\bibnamefont{{Thompson}}},
  \bibinfo{journal}{\apj} \textbf{\bibinfo{volume}{392}}, \bibinfo{pages}{L9}
  (\bibinfo{year}{1992}).

\bibitem[{\citenamefont{{Thompson} and {Duncan}}(1995)}]{ThompsonDuncan1995}
\bibinfo{author}{\bibfnamefont{C.}~\bibnamefont{{Thompson}}} \bibnamefont{and}
  \bibinfo{author}{\bibfnamefont{R.~C.} \bibnamefont{{Duncan}}},
  \bibinfo{journal}{\mnras} \textbf{\bibinfo{volume}{275}},
  \bibinfo{pages}{255} (\bibinfo{year}{1995}).

\bibitem[{\citenamefont{{Cheng} et~al.}(1996)\citenamefont{{Cheng}, {Epstein},
  {Guyer}, and {Young}}}]{Cheng1996}
\bibinfo{author}{\bibfnamefont{B.}~\bibnamefont{{Cheng}}},
  \bibinfo{author}{\bibfnamefont{R.~I.} \bibnamefont{{Epstein}}},
  \bibinfo{author}{\bibfnamefont{R.~A.} \bibnamefont{{Guyer}}},
  \bibnamefont{and} \bibinfo{author}{\bibfnamefont{A.~C.}
  \bibnamefont{{Young}}}, \bibinfo{journal}{Nature}
  \textbf{\bibinfo{volume}{382}}, \bibinfo{pages}{518} (\bibinfo{year}{1996}).

\bibitem[{\citenamefont{{Palmer}}(1999)}]{Palmer1999}
\bibinfo{author}{\bibfnamefont{D.~M.} \bibnamefont{{Palmer}}},
  \bibinfo{journal}{\apj} \textbf{\bibinfo{volume}{512}}, \bibinfo{pages}{L113}
  (\bibinfo{year}{1999}).

\bibitem[{\citenamefont{{Dubath} et~al.}(2005)\citenamefont{{Dubath}, {Foffa},
  {Gasparini}, {Maggiore}, and {Sturani}}}]{Dubath2005}
\bibinfo{author}{\bibfnamefont{F.}~\bibnamefont{{Dubath}}},
  \bibinfo{author}{\bibfnamefont{S.}~\bibnamefont{{Foffa}}},
  \bibinfo{author}{\bibfnamefont{M.~A.} \bibnamefont{{Gasparini}}},
  \bibinfo{author}{\bibfnamefont{M.}~\bibnamefont{{Maggiore}}},
  \bibnamefont{and}
  \bibinfo{author}{\bibfnamefont{R.}~\bibnamefont{{Sturani}}},
  \bibinfo{journal}{\prd} \textbf{\bibinfo{volume}{71}},
  \bibinfo{pages}{124003} (\bibinfo{year}{2005}).

\bibitem[{\citenamefont{Perna and Pons}(2011)}]{Perna:2011gt}
\bibinfo{author}{\bibfnamefont{R.}~\bibnamefont{Perna}} \bibnamefont{and}
  \bibinfo{author}{\bibfnamefont{J.~A.} \bibnamefont{Pons}}
  (\bibinfo{year}{2011}), \eprint{arXiv:1101.1098}.

\bibitem[{\citenamefont{Israel et~al.}(2005)}]{Israel:2005av}
\bibinfo{author}{\bibfnamefont{G.}~\bibnamefont{Israel}} \bibnamefont{et~al.},
  \bibinfo{journal}{Astrophys. J.} \textbf{\bibinfo{volume}{628}},
  \bibinfo{pages}{L53} (\bibinfo{year}{2005}).

\bibitem[{\citenamefont{Strohmayer and Watts}(2005)}]{Strohmayer:2005ks}
\bibinfo{author}{\bibfnamefont{T.~E.} \bibnamefont{Strohmayer}}
  \bibnamefont{and} \bibinfo{author}{\bibfnamefont{A.~L.} \bibnamefont{Watts}},
  \bibinfo{journal}{Astrophys. J.} \textbf{\bibinfo{volume}{632}},
  \bibinfo{pages}{L111} (\bibinfo{year}{2005}).

\bibitem[{\citenamefont{Levin}(2007)}]{Levin:2006qd}
\bibinfo{author}{\bibfnamefont{Y.}~\bibnamefont{Levin}}, \bibinfo{journal}{Mon.
  Not. Roy. Astron. Soc.} \textbf{\bibinfo{volume}{377}}, \bibinfo{pages}{159}
  (\bibinfo{year}{2007}).

\bibitem[{\citenamefont{Lyutikov}(2006)}]{Lyutikov:2005un}
\bibinfo{author}{\bibfnamefont{M.}~\bibnamefont{Lyutikov}},
  \bibinfo{journal}{Mon. Not. Roy. Astron. Soc.}
  \textbf{\bibinfo{volume}{367}}, \bibinfo{pages}{1594} (\bibinfo{year}{2006}).

\bibitem[{\citenamefont{van Hoven and Levin}(2010)}]{vanHoven:2010gy}
\bibinfo{author}{\bibfnamefont{M.}~\bibnamefont{van Hoven}} \bibnamefont{and}
  \bibinfo{author}{\bibfnamefont{Y.}~\bibnamefont{Levin}}
  (\bibinfo{year}{2010}), \eprint{arXiv:1006.0348}.

\bibitem[{\citenamefont{Gabler et~al.}(2010)\citenamefont{Gabler, Cerda-Duran,
  Font, Muller, and Stergioulas}}]{Gabler:2010rp}
\bibinfo{author}{\bibfnamefont{M.}~\bibnamefont{Gabler}},
  \bibinfo{author}{\bibfnamefont{P.}~\bibnamefont{Cerda-Duran}},
  \bibinfo{author}{\bibfnamefont{J.~A.} \bibnamefont{Font}},
  \bibinfo{author}{\bibfnamefont{E.}~\bibnamefont{Muller}}, \bibnamefont{and}
  \bibinfo{author}{\bibfnamefont{N.}~\bibnamefont{Stergioulas}}
  (\bibinfo{year}{2010}), \eprint{arXiv:1007.0856}.

\bibitem[{\citenamefont{{Lindblom} and {Detweiler}}(1983)}]{Detweiler1983}
\bibinfo{author}{\bibfnamefont{L.}~\bibnamefont{{Lindblom}}} \bibnamefont{and}
  \bibinfo{author}{\bibfnamefont{S.~L.} \bibnamefont{{Detweiler}}},
  \bibinfo{journal}{\apj Suppl.} \textbf{\bibinfo{volume}{53}},
  \bibinfo{pages}{73} (\bibinfo{year}{1983}).

\bibitem[{\citenamefont{{de Freitas Pacheco}}(1998)}]{Pacheco1998}
\bibinfo{author}{\bibfnamefont{J.~A.} \bibnamefont{{de Freitas Pacheco}}},
  \bibinfo{journal}{\aa} \textbf{\bibinfo{volume}{336}}, \bibinfo{pages}{397}
  (\bibinfo{year}{1998}).

\bibitem[{\citenamefont{{Gualtieri} et~al.}(2004)\citenamefont{{Gualtieri},
  {Pons}, {Miniutti}, and {G.}}}]{Gualtieri2004}
\bibinfo{author}{\bibfnamefont{L.}~\bibnamefont{{Gualtieri}}},
  \bibinfo{author}{\bibfnamefont{J.~A.} \bibnamefont{{Pons}}},
  \bibinfo{author}{\bibnamefont{{Miniutti}}}, \bibnamefont{and}
  \bibinfo{author}{\bibnamefont{{G.}}}, \bibinfo{journal}{Phys. Rev. D}
  \textbf{\bibinfo{volume}{70}}, \bibinfo{pages}{084009}
  (\bibinfo{year}{2004}).

\bibitem[{\citenamefont{Hurley et~al.}(2005)}]{Hurley:2005zs}
\bibinfo{author}{\bibfnamefont{K.}~\bibnamefont{Hurley}} \bibnamefont{et~al.},
  \bibinfo{journal}{Nature} \textbf{\bibinfo{volume}{434}},
  \bibinfo{pages}{1098} (\bibinfo{year}{2005}).

\bibitem[{\citenamefont{{Blaes} et~al.}(1989)\citenamefont{{Blaes},
  {Blandford}, {Goldreich}, and {Madau}}}]{Blaes1989}
\bibinfo{author}{\bibfnamefont{O.}~\bibnamefont{{Blaes}}},
  \bibinfo{author}{\bibfnamefont{R.}~\bibnamefont{{Blandford}}},
  \bibinfo{author}{\bibfnamefont{P.}~\bibnamefont{{Goldreich}}},
  \bibnamefont{and} \bibinfo{author}{\bibfnamefont{P.}~\bibnamefont{{Madau}}},
  \bibinfo{journal}{\apj} \textbf{\bibinfo{volume}{343}}, \bibinfo{pages}{839}
  (\bibinfo{year}{1989}).

\bibitem[{\citenamefont{Horvath}(2005)}]{Horvath:2005ta}
\bibinfo{author}{\bibfnamefont{J.~E.} \bibnamefont{Horvath}},
  \bibinfo{journal}{Mod. Phys. Lett. A} \textbf{\bibinfo{volume}{20}},
  \bibinfo{pages}{2799} (\bibinfo{year}{2005}).

\bibitem[{\citenamefont{Owen}(2005)}]{Owen:2005fn}
\bibinfo{author}{\bibfnamefont{B.~J.} \bibnamefont{Owen}},
  \bibinfo{journal}{Phys. Rev. Lett.} \textbf{\bibinfo{volume}{95}},
  \bibinfo{pages}{211101} (\bibinfo{year}{2005}).

\bibitem[{\citenamefont{Xu et~al.}(2006)\citenamefont{Xu, Tao, and
  Yang}}]{Xu:2006mp}
\bibinfo{author}{\bibfnamefont{R.-X.} \bibnamefont{Xu}},
  \bibinfo{author}{\bibfnamefont{D.~J.} \bibnamefont{Tao}}, \bibnamefont{and}
  \bibinfo{author}{\bibfnamefont{Y.}~\bibnamefont{Yang}},
  \bibinfo{journal}{Mon. Not. Roy. Astron. Soc.}
  \textbf{\bibinfo{volume}{373}}, \bibinfo{pages}{L85} (\bibinfo{year}{2006}).

\bibitem[{\citenamefont{{Horowitz} and {Kadau}}(2009)}]{Horowitz:2009ya}
\bibinfo{author}{\bibfnamefont{C.~J.} \bibnamefont{{Horowitz}}}
  \bibnamefont{and} \bibinfo{author}{\bibfnamefont{K.}~\bibnamefont{{Kadau}}},
  \bibinfo{journal}{Phys. Rev. Lett.} \textbf{\bibinfo{volume}{102}},
  \bibinfo{pages}{191102} (\bibinfo{year}{2009}).

\bibitem[{\citenamefont{{Xu}}(2003)}]{Xu2003}
\bibinfo{author}{\bibfnamefont{R.~X.} \bibnamefont{{Xu}}}, in
  \emph{\bibinfo{booktitle}{High Energy Processes and Phenomena in
  Astrophysics}}, edited by \bibinfo{editor}{\bibnamefont{{X.~D.~Li,
  V.~Trimble, \& Z.~R.~Wang}}} (\bibinfo{year}{2003}), vol.
  \bibinfo{volume}{214} of \emph{\bibinfo{series}{IAU Symposium}}, pp.
  \bibinfo{pages}{191--+}.

\bibitem[{\citenamefont{{Mannarelli} et~al.}(2007)\citenamefont{{Mannarelli},
  {Rajagopal}, and {Sharma}}}]{Mannarelli2007}
\bibinfo{author}{\bibfnamefont{M.}~\bibnamefont{{Mannarelli}}},
  \bibinfo{author}{\bibfnamefont{K.}~\bibnamefont{{Rajagopal}}},
  \bibnamefont{and} \bibinfo{author}{\bibfnamefont{R.}~\bibnamefont{{Sharma}}},
  \bibinfo{journal}{Phys. Rev. D} \textbf{\bibinfo{volume}{76}},
  \bibinfo{pages}{074026} (\bibinfo{year}{2007}).

\bibitem[{\citenamefont{Johnson-McDaniel and Owen}(2011)}]{exotic}
\bibinfo{author}{\bibfnamefont{N.~K.} \bibnamefont{Johnson-McDaniel}}
  \bibnamefont{and} \bibinfo{author}{\bibfnamefont{B.~J.} \bibnamefont{Owen}}
  (\bibinfo{year}{2011}), \bibinfo{note}{in preparation}.

\bibitem[{\citenamefont{{Ioka}}(2001)}]{Ioka2001}
\bibinfo{author}{\bibfnamefont{K.}~\bibnamefont{{Ioka}}},
  \bibinfo{journal}{\mnras} \textbf{\bibinfo{volume}{327}},
  \bibinfo{pages}{639} (\bibinfo{year}{2001}).

\bibitem[{\citenamefont{Ott}(2009)}]{Ott:2008wt}
\bibinfo{author}{\bibfnamefont{C.~D.} \bibnamefont{Ott}},
  \bibinfo{journal}{Class. Quant. Grav.} \textbf{\bibinfo{volume}{26}},
  \bibinfo{pages}{063001} (\bibinfo{year}{2009}).

\bibitem[{\citenamefont{{Thompson} and {Duncan}}(2001)}]{ThompsonDuncan2001}
\bibinfo{author}{\bibfnamefont{C.}~\bibnamefont{{Thompson}}} \bibnamefont{and}
  \bibinfo{author}{\bibfnamefont{R.~C.} \bibnamefont{{Duncan}}},
  \bibinfo{journal}{\apj} \textbf{\bibinfo{volume}{561}}, \bibinfo{pages}{980}
  (\bibinfo{year}{2001}).

\bibitem[{\citenamefont{{Thompson} et~al.}(2002)\citenamefont{{Thompson},
  {Lyutikov}, and {Kulkarni}}}]{Thompson2002}
\bibinfo{author}{\bibfnamefont{C.}~\bibnamefont{{Thompson}}},
  \bibinfo{author}{\bibfnamefont{M.}~\bibnamefont{{Lyutikov}}},
  \bibnamefont{and} \bibinfo{author}{\bibfnamefont{S.~R.}
  \bibnamefont{{Kulkarni}}}, \bibinfo{journal}{\apj}
  \textbf{\bibinfo{volume}{574}}, \bibinfo{pages}{332} (\bibinfo{year}{2002}).

\bibitem[{\citenamefont{{Thompson} and {Duncan}}(1993)}]{ThompsonDuncan1993}
\bibinfo{author}{\bibfnamefont{C.}~\bibnamefont{{Thompson}}} \bibnamefont{and}
  \bibinfo{author}{\bibfnamefont{R.~C.} \bibnamefont{{Duncan}}},
  \bibinfo{journal}{\apj} \textbf{\bibinfo{volume}{408}}, \bibinfo{pages}{194}
  (\bibinfo{year}{1993}).

\bibitem[{\citenamefont{{Usov}}(1992)}]{Usov1992}
\bibinfo{author}{\bibfnamefont{V.~V.} \bibnamefont{{Usov}}},
  \bibinfo{journal}{Nature} \textbf{\bibinfo{volume}{357}},
  \bibinfo{pages}{472} (\bibinfo{year}{1992}).

\bibitem[{\citenamefont{{Klu{\'z}niak} and {Ruderman}}(1998)}]{Klu1998}
\bibinfo{author}{\bibfnamefont{W.}~\bibnamefont{{Klu{\'z}niak}}}
  \bibnamefont{and}
  \bibinfo{author}{\bibfnamefont{M.}~\bibnamefont{{Ruderman}}},
  \bibinfo{journal}{\apj} \textbf{\bibinfo{volume}{505}}, \bibinfo{pages}{L113}
  (\bibinfo{year}{1998}).

\bibitem[{\citenamefont{{Wheeler} et~al.}(2000)\citenamefont{{Wheeler}, {Yi},
  {H{\"o}flich}, and {Wang}}}]{Wheeler2000}
\bibinfo{author}{\bibfnamefont{J.~C.} \bibnamefont{{Wheeler}}},
  \bibinfo{author}{\bibfnamefont{I.}~\bibnamefont{{Yi}}},
  \bibinfo{author}{\bibfnamefont{P.}~\bibnamefont{{H{\"o}flich}}},
  \bibnamefont{and} \bibinfo{author}{\bibfnamefont{L.}~\bibnamefont{{Wang}}},
  \bibinfo{journal}{\apj} \textbf{\bibinfo{volume}{537}}, \bibinfo{pages}{810}
  (\bibinfo{year}{2000}).

\bibitem[{\citenamefont{Kaminker et~al.}(2006)}]{KaminkerEtAl2006}
\bibinfo{author}{\bibfnamefont{A.~D.} \bibnamefont{Kaminker}}
  \bibnamefont{et~al.}, \bibinfo{journal}{Mon. Not. Roy. Astron. Soc.}
  \textbf{\bibinfo{volume}{371}}, \bibinfo{pages}{477} (\bibinfo{year}{2006}).

\bibitem[{\citenamefont{{Woods} et~al.}(2007)\citenamefont{{Woods},
  {Kouveliotou}, {Finger}, {G{\"o}{\u g}{\"u}{\c s}}, {Wilson}, {Patel},
  {Hurley}, and {Swank}}}]{WoodsEtAl2007}
\bibinfo{author}{\bibfnamefont{P.~M.} \bibnamefont{{Woods}}},
  \bibinfo{author}{\bibfnamefont{C.}~\bibnamefont{{Kouveliotou}}},
  \bibinfo{author}{\bibfnamefont{M.~H.} \bibnamefont{{Finger}}},
  \bibinfo{author}{\bibfnamefont{E.}~\bibnamefont{{G{\"o}{\u g}{\"u}{\c s}}}},
  \bibinfo{author}{\bibfnamefont{C.~A.} \bibnamefont{{Wilson}}},
  \bibinfo{author}{\bibfnamefont{S.~K.} \bibnamefont{{Patel}}},
  \bibinfo{author}{\bibfnamefont{K.}~\bibnamefont{{Hurley}}}, \bibnamefont{and}
  \bibinfo{author}{\bibfnamefont{J.~H.} \bibnamefont{{Swank}}},
  \bibinfo{journal}{\apj} \textbf{\bibinfo{volume}{654}}, \bibinfo{pages}{470}
  (\bibinfo{year}{2007}).

\bibitem[{\citenamefont{{Mereghetti}}(2008)}]{Mereghetti2008}
\bibinfo{author}{\bibfnamefont{S.}~\bibnamefont{{Mereghetti}}},
  \bibinfo{journal}{\aa Rev.} \textbf{\bibinfo{volume}{15}},
  \bibinfo{pages}{225} (\bibinfo{year}{2008}).

\bibitem[{\citenamefont{{Lander} and {Jones}}(2009)}]{Lander2009}
\bibinfo{author}{\bibfnamefont{S.~K.} \bibnamefont{{Lander}}} \bibnamefont{and}
  \bibinfo{author}{\bibfnamefont{D.~I.} \bibnamefont{{Jones}}},
  \bibinfo{journal}{\mnras} \textbf{\bibinfo{volume}{395}},
  \bibinfo{pages}{2162} (\bibinfo{year}{2009}).

\bibitem[{\citenamefont{{Chandrasekhar}}(1933)}]{Chandrasekhar1933}
\bibinfo{author}{\bibfnamefont{S.}~\bibnamefont{{Chandrasekhar}}},
  \bibinfo{journal}{\mnras} \textbf{\bibinfo{volume}{93}}, \bibinfo{pages}{390}
  (\bibinfo{year}{1933}).

\bibitem[{\citenamefont{{Chandrasekhar} and
  {Lebovitz}}(1962)}]{Chandrasekhar1962}
\bibinfo{author}{\bibfnamefont{S.}~\bibnamefont{{Chandrasekhar}}}
  \bibnamefont{and} \bibinfo{author}{\bibfnamefont{N.~R.}
  \bibnamefont{{Lebovitz}}}, \bibinfo{journal}{\apj}
  \textbf{\bibinfo{volume}{136}}, \bibinfo{pages}{1082} (\bibinfo{year}{1962}).

\bibitem[{\citenamefont{{Chandrasekhar} and {Fermi}}(1953)}]{Chandrasekhar1953}
\bibinfo{author}{\bibfnamefont{S.}~\bibnamefont{{Chandrasekhar}}}
  \bibnamefont{and} \bibinfo{author}{\bibfnamefont{E.}~\bibnamefont{{Fermi}}},
  \bibinfo{journal}{\apj} \textbf{\bibinfo{volume}{118}}, \bibinfo{pages}{116}
  (\bibinfo{year}{1953}).

\bibitem[{\citenamefont{{Monaghan}}(1965)}]{Monaghan1965}
\bibinfo{author}{\bibfnamefont{J.~J.} \bibnamefont{{Monaghan}}},
  \bibinfo{journal}{\mnras} \textbf{\bibinfo{volume}{131}},
  \bibinfo{pages}{105} (\bibinfo{year}{1965}).

\bibitem[{\citenamefont{{Monaghan}}(1966{\natexlab{a}})}]{Monaghan1966a}
\bibinfo{author}{\bibfnamefont{F.~F.} \bibnamefont{{Monaghan}}},
  \bibinfo{journal}{\mnras} \textbf{\bibinfo{volume}{132}}, \bibinfo{pages}{1}
  (\bibinfo{year}{1966}{\natexlab{a}}).

\bibitem[{\citenamefont{{Monaghan}}(1966{\natexlab{b}})}]{Monaghan1966b}
\bibinfo{author}{\bibfnamefont{J.~J.} \bibnamefont{{Monaghan}}},
  \bibinfo{journal}{\mnras} \textbf{\bibinfo{volume}{134}},
  \bibinfo{pages}{275} (\bibinfo{year}{1966}{\natexlab{b}}).

\bibitem[{\citenamefont{{Roxburgh}}(1966)}]{Roxburgh1966}
\bibinfo{author}{\bibfnamefont{I.~W.} \bibnamefont{{Roxburgh}}},
  \bibinfo{journal}{\mnras} \textbf{\bibinfo{volume}{132}},
  \bibinfo{pages}{347} (\bibinfo{year}{1966}).

\bibitem[{\citenamefont{{Trehan} and {Billings}}(1971)}]{Trehan1971}
\bibinfo{author}{\bibfnamefont{S.~K.} \bibnamefont{{Trehan}}} \bibnamefont{and}
  \bibinfo{author}{\bibfnamefont{D.~F.} \bibnamefont{{Billings}}},
  \bibinfo{journal}{\apj} \textbf{\bibinfo{volume}{169}}, \bibinfo{pages}{567}
  (\bibinfo{year}{1971}).

\bibitem[{\citenamefont{{Trehan} and {Uberoi}}(1972)}]{Trehan1972}
\bibinfo{author}{\bibfnamefont{S.~K.} \bibnamefont{{Trehan}}} \bibnamefont{and}
  \bibinfo{author}{\bibfnamefont{M.~S.} \bibnamefont{{Uberoi}}},
  \bibinfo{journal}{\apj} \textbf{\bibinfo{volume}{175}}, \bibinfo{pages}{161}
  (\bibinfo{year}{1972}).

\bibitem[{\citenamefont{{Reisenegger}}(2009)}]{Reisenegger2009}
\bibinfo{author}{\bibfnamefont{A.}~\bibnamefont{{Reisenegger}}},
  \bibinfo{journal}{\aa} \textbf{\bibinfo{volume}{499}}, \bibinfo{pages}{557}
  (\bibinfo{year}{2009}).

\bibitem[{\citenamefont{{Ioka} and {Sasaki}}(2004)}]{Ioka2004}
\bibinfo{author}{\bibfnamefont{K.}~\bibnamefont{{Ioka}}} \bibnamefont{and}
  \bibinfo{author}{\bibfnamefont{M.}~\bibnamefont{{Sasaki}}},
  \bibinfo{journal}{\apj} \textbf{\bibinfo{volume}{600}}, \bibinfo{pages}{296}
  (\bibinfo{year}{2004}).

\bibitem[{\citenamefont{{Ioka} and {Sasaki}}(2003)}]{Ioka2003}
\bibinfo{author}{\bibfnamefont{K.}~\bibnamefont{{Ioka}}} \bibnamefont{and}
  \bibinfo{author}{\bibfnamefont{M.}~\bibnamefont{{Sasaki}}},
  \bibinfo{journal}{Phys. Rev. D} \textbf{\bibinfo{volume}{67}},
  \bibinfo{pages}{124026} (\bibinfo{year}{2003}).

\bibitem[{\citenamefont{{Colaiuda} et~al.}(2008)\citenamefont{{Colaiuda},
  {Ferrari}, {Gualtieri}, and {Pons}}}]{Colaiuda2008}
\bibinfo{author}{\bibfnamefont{A.}~\bibnamefont{{Colaiuda}}},
  \bibinfo{author}{\bibfnamefont{V.}~\bibnamefont{{Ferrari}}},
  \bibinfo{author}{\bibfnamefont{L.}~\bibnamefont{{Gualtieri}}},
  \bibnamefont{and} \bibinfo{author}{\bibfnamefont{J.~A.}
  \bibnamefont{{Pons}}}, \bibinfo{journal}{\mnras}
  \textbf{\bibinfo{volume}{385}}, \bibinfo{pages}{2080} (\bibinfo{year}{2008}).

\bibitem[{\citenamefont{Ciolfi et~al.}(2009)\citenamefont{Ciolfi, Ferrari,
  Gualtieri, and Pons}}]{Ciolfi2009}
\bibinfo{author}{\bibfnamefont{R.}~\bibnamefont{Ciolfi}},
  \bibinfo{author}{\bibfnamefont{V.}~\bibnamefont{Ferrari}},
  \bibinfo{author}{\bibfnamefont{L.}~\bibnamefont{Gualtieri}},
  \bibnamefont{and} \bibinfo{author}{\bibfnamefont{J.~A.} \bibnamefont{Pons}},
  \bibinfo{journal}{Mon. Not. Roy. Astron. Soc.}
  \textbf{\bibinfo{volume}{397}}, \bibinfo{pages}{913} (\bibinfo{year}{2009}).

\bibitem[{\citenamefont{Ciolfi et~al.}(2010)\citenamefont{Ciolfi, Ferrari, and
  Gualtieri}}]{Ciolfi2010}
\bibinfo{author}{\bibfnamefont{R.}~\bibnamefont{Ciolfi}},
  \bibinfo{author}{\bibfnamefont{V.}~\bibnamefont{Ferrari}}, \bibnamefont{and}
  \bibinfo{author}{\bibfnamefont{L.}~\bibnamefont{Gualtieri}},
  \bibinfo{journal}{Mon. Not. Roy. Astron. Soc.}
  \textbf{\bibinfo{volume}{406}}, \bibinfo{pages}{2540} (\bibinfo{year}{2010}).

\bibitem[{\citenamefont{{Rea} et~al.}(2010)\citenamefont{{Rea}, {Esposito},
  {Turolla}, {Israel}, {Zane}, {Stella}, {Mereghetti}, {Tiengo}, {G{\"o}tz},
  {G{\"o}{\u g}{\"u}{\c s}} et~al.}}]{ReaEtAl2010}
\bibinfo{author}{\bibfnamefont{N.}~\bibnamefont{{Rea}}},
  \bibinfo{author}{\bibfnamefont{P.}~\bibnamefont{{Esposito}}},
  \bibinfo{author}{\bibfnamefont{R.}~\bibnamefont{{Turolla}}},
  \bibinfo{author}{\bibfnamefont{G.~L.} \bibnamefont{{Israel}}},
  \bibinfo{author}{\bibfnamefont{S.}~\bibnamefont{{Zane}}},
  \bibinfo{author}{\bibfnamefont{L.}~\bibnamefont{{Stella}}},
  \bibinfo{author}{\bibfnamefont{S.}~\bibnamefont{{Mereghetti}}},
  \bibinfo{author}{\bibfnamefont{A.}~\bibnamefont{{Tiengo}}},
  \bibinfo{author}{\bibfnamefont{D.}~\bibnamefont{{G{\"o}tz}}},
  \bibinfo{author}{\bibfnamefont{E.}~\bibnamefont{{G{\"o}{\u g}{\"u}{\c s}}}},
  \bibnamefont{et~al.}, \bibinfo{journal}{Science}
  \textbf{\bibinfo{volume}{330}}, \bibinfo{pages}{944} (\bibinfo{year}{2010}).

\bibitem[{\citenamefont{{Haskell} et~al.}(2008)\citenamefont{{Haskell},
  {Samuelsson}, {Glampedakis}, and {Andersson}}}]{Haskell2008}
\bibinfo{author}{\bibfnamefont{B.}~\bibnamefont{{Haskell}}},
  \bibinfo{author}{\bibfnamefont{L.}~\bibnamefont{{Samuelsson}}},
  \bibinfo{author}{\bibfnamefont{K.}~\bibnamefont{{Glampedakis}}},
  \bibnamefont{and}
  \bibinfo{author}{\bibfnamefont{N.}~\bibnamefont{{Andersson}}},
  \bibinfo{journal}{\mnras} \textbf{\bibinfo{volume}{385}},
  \bibinfo{pages}{531} (\bibinfo{year}{2008}).

\bibitem[{\citenamefont{{Braithwaite}}(2009)}]{Braithwaite2008}
\bibinfo{author}{\bibfnamefont{J.}~\bibnamefont{{Braithwaite}}},
  \bibinfo{journal}{\mnras} \textbf{\bibinfo{volume}{397}},
  \bibinfo{pages}{763} (\bibinfo{year}{2009}).

\bibitem[{\citenamefont{{Tomimura} and {Eriguchi}}(2005)}]{Tomimura2005}
\bibinfo{author}{\bibfnamefont{Y.}~\bibnamefont{{Tomimura}}} \bibnamefont{and}
  \bibinfo{author}{\bibfnamefont{Y.}~\bibnamefont{{Eriguchi}}},
  \bibinfo{journal}{\mnras} \textbf{\bibinfo{volume}{359}},
  \bibinfo{pages}{1117} (\bibinfo{year}{2005}).

\bibitem[{\citenamefont{{Yoshida} and {Eriguchi}}(2006)}]{Yoshida2006a}
\bibinfo{author}{\bibfnamefont{S.}~\bibnamefont{{Yoshida}}} \bibnamefont{and}
  \bibinfo{author}{\bibfnamefont{Y.}~\bibnamefont{{Eriguchi}}},
  \bibinfo{journal}{\apj Suppl.} \textbf{\bibinfo{volume}{164}},
  \bibinfo{pages}{156} (\bibinfo{year}{2006}).

\bibitem[{\citenamefont{{Yoshida} et~al.}(2006)\citenamefont{{Yoshida},
  {Yoshida}, and {Eriguchi}}}]{Yoshida2006b}
\bibinfo{author}{\bibfnamefont{S.}~\bibnamefont{{Yoshida}}},
  \bibinfo{author}{\bibfnamefont{S.}~\bibnamefont{{Yoshida}}},
  \bibnamefont{and}
  \bibinfo{author}{\bibfnamefont{Y.}~\bibnamefont{{Eriguchi}}},
  \bibinfo{journal}{\apj} \textbf{\bibinfo{volume}{651}}, \bibinfo{pages}{462}
  (\bibinfo{year}{2006}).

\bibitem[{\citenamefont{{Prendergast}}(1956)}]{Prendergast1956}
\bibinfo{author}{\bibfnamefont{K.~H.} \bibnamefont{{Prendergast}}},
  \bibinfo{journal}{\apj} \textbf{\bibinfo{volume}{123}}, \bibinfo{pages}{498}
  (\bibinfo{year}{1956}).

\bibitem[{\citenamefont{{Markey} and {Tayler}}(1973)}]{Markey1973}
\bibinfo{author}{\bibfnamefont{P.}~\bibnamefont{{Markey}}} \bibnamefont{and}
  \bibinfo{author}{\bibfnamefont{R.~J.} \bibnamefont{{Tayler}}},
  \bibinfo{journal}{\mnras} \textbf{\bibinfo{volume}{163}}, \bibinfo{pages}{77}
  (\bibinfo{year}{1973}).

\bibitem[{\citenamefont{{Tayler}}(1973)}]{Tayler1973}
\bibinfo{author}{\bibfnamefont{R.~J.} \bibnamefont{{Tayler}}},
  \bibinfo{journal}{\mnras} \textbf{\bibinfo{volume}{161}},
  \bibinfo{pages}{365} (\bibinfo{year}{1973}).

\bibitem[{\citenamefont{{Wright}}(1973)}]{Wright1973}
\bibinfo{author}{\bibfnamefont{G.~A.~E.} \bibnamefont{{Wright}}},
  \bibinfo{journal}{\mnras} \textbf{\bibinfo{volume}{162}},
  \bibinfo{pages}{339} (\bibinfo{year}{1973}).

\bibitem[{\citenamefont{{Markey} and {Tayler}}(1974)}]{Markey1974}
\bibinfo{author}{\bibfnamefont{P.}~\bibnamefont{{Markey}}} \bibnamefont{and}
  \bibinfo{author}{\bibfnamefont{R.~J.} \bibnamefont{{Tayler}}},
  \bibinfo{journal}{\mnras} \textbf{\bibinfo{volume}{168}},
  \bibinfo{pages}{505} (\bibinfo{year}{1974}).

\bibitem[{\citenamefont{{Spruit}}(1998)}]{Spruit1998}
\bibinfo{author}{\bibfnamefont{H.~C.} \bibnamefont{{Spruit}}},
  \bibinfo{journal}{\aa} \textbf{\bibinfo{volume}{333}}, \bibinfo{pages}{603}
  (\bibinfo{year}{1998}).

\bibitem[{\citenamefont{{Braithwaite} and {Nordlund}}(2006)}]{Braithwaite2006a}
\bibinfo{author}{\bibfnamefont{J.}~\bibnamefont{{Braithwaite}}}
  \bibnamefont{and}
  \bibinfo{author}{\bibfnamefont{{\AA}.}~\bibnamefont{{Nordlund}}},
  \bibinfo{journal}{\aa} \textbf{\bibinfo{volume}{450}}, \bibinfo{pages}{1077}
  (\bibinfo{year}{2006}).

\bibitem[{\citenamefont{{Braithwaite}}(2006{\natexlab{a}})}]{Braithwaite2006b}
\bibinfo{author}{\bibfnamefont{J.}~\bibnamefont{{Braithwaite}}},
  \bibinfo{journal}{\aa} \textbf{\bibinfo{volume}{453}}, \bibinfo{pages}{687}
  (\bibinfo{year}{2006}{\natexlab{a}}).

\bibitem[{\citenamefont{{Braithwaite}}(2006{\natexlab{b}})}]{Braithwaite2006c}
\bibinfo{author}{\bibfnamefont{J.}~\bibnamefont{{Braithwaite}}},
  \bibinfo{journal}{\aa} \textbf{\bibinfo{volume}{449}}, \bibinfo{pages}{451}
  (\bibinfo{year}{2006}{\natexlab{b}}).

\bibitem[{\citenamefont{{Bonanno} and {Urpin}}(2008)}]{Bonanno2008}
\bibinfo{author}{\bibfnamefont{A.}~\bibnamefont{{Bonanno}}} \bibnamefont{and}
  \bibinfo{author}{\bibfnamefont{V.}~\bibnamefont{{Urpin}}},
  \bibinfo{journal}{\aa} \textbf{\bibinfo{volume}{477}}, \bibinfo{pages}{35}
  (\bibinfo{year}{2008}).

\bibitem[{\citenamefont{{Braithwaite} and {Spruit}}(2004)}]{Braithwaite2004}
\bibinfo{author}{\bibfnamefont{J.}~\bibnamefont{{Braithwaite}}}
  \bibnamefont{and} \bibinfo{author}{\bibfnamefont{H.~C.}
  \bibnamefont{{Spruit}}}, \bibinfo{journal}{Nature}
  \textbf{\bibinfo{volume}{431}}, \bibinfo{pages}{819} (\bibinfo{year}{2004}).

\bibitem[{\citenamefont{Akgun and Wasserman}(2008)}]{Akgun2008}
\bibinfo{author}{\bibfnamefont{T.}~\bibnamefont{Akgun}} \bibnamefont{and}
  \bibinfo{author}{\bibfnamefont{I.}~\bibnamefont{Wasserman}},
  \bibinfo{journal}{Mon. Not. Roy. Astron. Soc.}
  \textbf{\bibinfo{volume}{383}}, \bibinfo{pages}{1551} (\bibinfo{year}{2008}).

\bibitem[{\citenamefont{{Kiuchi} and {Kotake}}(2008)}]{Kiuchi2008a}
\bibinfo{author}{\bibfnamefont{K.}~\bibnamefont{{Kiuchi}}} \bibnamefont{and}
  \bibinfo{author}{\bibfnamefont{K.}~\bibnamefont{{Kotake}}},
  \bibinfo{journal}{\mnras} \textbf{\bibinfo{volume}{385}},
  \bibinfo{pages}{1327} (\bibinfo{year}{2008}).

\bibitem[{\citenamefont{{Douchin} and {Haensel}}(2001)}]{Douchin2001}
\bibinfo{author}{\bibfnamefont{F.}~\bibnamefont{{Douchin}}} \bibnamefont{and}
  \bibinfo{author}{\bibfnamefont{P.}~\bibnamefont{{Haensel}}},
  \bibinfo{journal}{\aa} \textbf{\bibinfo{volume}{380}}, \bibinfo{pages}{151}
  (\bibinfo{year}{2001}).

\bibitem[{\citenamefont{{Pandharipande} and
  {Ravenhall}}(1989)}]{Pandharipande1989}
\bibinfo{author}{\bibfnamefont{V.~R.} \bibnamefont{{Pandharipande}}}
  \bibnamefont{and} \bibinfo{author}{\bibfnamefont{D.~G.}
  \bibnamefont{{Ravenhall}}}, \bibinfo{journal}{Proc. 205: Nuclear Matter \&
  Heavy Ion Collisions} p. \bibinfo{pages}{103} (\bibinfo{year}{1989}).

\bibitem[{\citenamefont{{Shen} et~al.}(1998)\citenamefont{{Shen}, {Toki},
  {Oyamatsu}, and {Sumiyoshi}}}]{Shen1998}
\bibinfo{author}{\bibfnamefont{H.}~\bibnamefont{{Shen}}},
  \bibinfo{author}{\bibfnamefont{H.}~\bibnamefont{{Toki}}},
  \bibinfo{author}{\bibfnamefont{K.}~\bibnamefont{{Oyamatsu}}},
  \bibnamefont{and}
  \bibinfo{author}{\bibfnamefont{K.}~\bibnamefont{{Sumiyoshi}}},
  \bibinfo{journal}{Prog. Theor. Phys.} \textbf{\bibinfo{volume}{100}},
  \bibinfo{pages}{1013} (\bibinfo{year}{1998}).

\bibitem[{\citenamefont{{Lattimer} and {Douglas Swesty}}(1991)}]{Lattimer1991}
\bibinfo{author}{\bibfnamefont{J.~M.} \bibnamefont{{Lattimer}}}
  \bibnamefont{and} \bibinfo{author}{\bibfnamefont{F.}~\bibnamefont{{Douglas
  Swesty}}}, \bibinfo{journal}{Nucl. Phys. A} \textbf{\bibinfo{volume}{535}},
  \bibinfo{pages}{331} (\bibinfo{year}{1991}).

\bibitem[{\citenamefont{{Kiuchi} et~al.}(2009)\citenamefont{{Kiuchi}, {Kotake},
  and {Yoshida}}}]{Kiuchi2009}
\bibinfo{author}{\bibfnamefont{K.}~\bibnamefont{{Kiuchi}}},
  \bibinfo{author}{\bibfnamefont{K.}~\bibnamefont{{Kotake}}}, \bibnamefont{and}
  \bibinfo{author}{\bibfnamefont{S.}~\bibnamefont{{Yoshida}}},
  \bibinfo{journal}{\apj} \textbf{\bibinfo{volume}{698}}, \bibinfo{pages}{541}
  (\bibinfo{year}{2009}).

\bibitem[{\citenamefont{{Corsi} and {Owen}}(2009)}]{LIGOnote}
\bibinfo{author}{\bibfnamefont{A.}~\bibnamefont{{Corsi}}} \bibnamefont{and}
  \bibinfo{author}{\bibfnamefont{B.~J.} \bibnamefont{{Owen}}}
  (\bibinfo{year}{2009}), \bibinfo{note}{{Technical Document LIGO-T0900242 and
  VIR-NOT-ROM-028A-09}}, \urlprefix\url{http://dcc.ligo.org}.

\bibitem[{\citenamefont{{Chandrasekhar}}(1967)}]{Chandrasekhar1967}
\bibinfo{author}{\bibfnamefont{S.}~\bibnamefont{{Chandrasekhar}}},
  \emph{\bibinfo{title}{{An introduction to the study of stellar structure}}}
  (\bibinfo{publisher}{Dover}, \bibinfo{address}{New York},
  \bibinfo{year}{1967}).

\bibitem[{\citenamefont{{Woltjer}}(1959)}]{Woltjer1959}
\bibinfo{author}{\bibfnamefont{L.}~\bibnamefont{{Woltjer}}},
  \bibinfo{journal}{\apj} \textbf{\bibinfo{volume}{130}}, \bibinfo{pages}{405}
  (\bibinfo{year}{1959}).

\bibitem[{\citenamefont{{Chandrasekhar}}(1961)}]{Chandrasekhar1961}
\bibinfo{author}{\bibfnamefont{S.}~\bibnamefont{{Chandrasekhar}}},
  \emph{\bibinfo{title}{{Hydrodynamic and hydromagnetic stability}}},
  International Series of Monographs on Physics
  (\bibinfo{publisher}{Clarendon}, \bibinfo{address}{Oxford},
  \bibinfo{year}{1961}).

\bibitem[{\citenamefont{{Woltjer}}(1960)}]{Woltjer1960}
\bibinfo{author}{\bibfnamefont{L.}~\bibnamefont{{Woltjer}}},
  \bibinfo{journal}{\apj} \textbf{\bibinfo{volume}{131}}, \bibinfo{pages}{227}
  (\bibinfo{year}{1960}).

\bibitem[{\citenamefont{{Chandrasekhar} and
  {Prendergast}}(1956)}]{Chandrasekhar1956}
\bibinfo{author}{\bibfnamefont{S.}~\bibnamefont{{Chandrasekhar}}}
  \bibnamefont{and} \bibinfo{author}{\bibfnamefont{K.~H.}
  \bibnamefont{{Prendergast}}}, \bibinfo{journal}{Proc. Nat. Acad. Sci.}
  \textbf{\bibinfo{volume}{42}}, \bibinfo{pages}{5} (\bibinfo{year}{1956}).

\bibitem[{\citenamefont{{Ledoux}}(1946)}]{Ledoux1946}
\bibinfo{author}{\bibfnamefont{P.}~\bibnamefont{{Ledoux}}},
  \bibinfo{journal}{\apj} \textbf{\bibinfo{volume}{104}}, \bibinfo{pages}{333}
  (\bibinfo{year}{1946}).

\bibitem[{\citenamefont{{Chandrasekhar} and {Lebovitz}}(1968)}]{Chan1968}
\bibinfo{author}{\bibfnamefont{S.}~\bibnamefont{{Chandrasekhar}}}
  \bibnamefont{and} \bibinfo{author}{\bibfnamefont{N.~R.}
  \bibnamefont{{Lebovitz}}}, \bibinfo{journal}{\apj}
  \textbf{\bibinfo{volume}{152}}, \bibinfo{pages}{267} (\bibinfo{year}{1968}).

\bibitem[{\citenamefont{Read et~al.}(2009)\citenamefont{Read, Lackey, Owen, and
  Friedman}}]{Read2008}
\bibinfo{author}{\bibfnamefont{J.~S.} \bibnamefont{Read}},
  \bibinfo{author}{\bibfnamefont{B.~D.} \bibnamefont{Lackey}},
  \bibinfo{author}{\bibfnamefont{B.~J.} \bibnamefont{Owen}}, \bibnamefont{and}
  \bibinfo{author}{\bibfnamefont{J.~L.} \bibnamefont{Friedman}},
  \bibinfo{journal}{Phys. Rev. D} \textbf{\bibinfo{volume}{79}},
  \bibinfo{pages}{124032} (\bibinfo{year}{2009}).

\bibitem[{\citenamefont{Demorest et~al.}(2010)\citenamefont{Demorest, Pennucci,
  Ransom, Roberts, and Hessels}}]{Demorest:2010bx}
\bibinfo{author}{\bibfnamefont{P.}~\bibnamefont{Demorest}},
  \bibinfo{author}{\bibfnamefont{T.}~\bibnamefont{Pennucci}},
  \bibinfo{author}{\bibfnamefont{S.}~\bibnamefont{Ransom}},
  \bibinfo{author}{\bibfnamefont{M.}~\bibnamefont{Roberts}}, \bibnamefont{and}
  \bibinfo{author}{\bibfnamefont{J.}~\bibnamefont{Hessels}},
  \bibinfo{journal}{Nature} \textbf{\bibinfo{volume}{467}},
  \bibinfo{pages}{1081} (\bibinfo{year}{2010}).

\bibitem[{\citenamefont{{Pavan} et~al.}(2009)\citenamefont{{Pavan}, {Turolla},
  {Zane}, and {Nobili}}}]{Pavan2009}
\bibinfo{author}{\bibfnamefont{L.}~\bibnamefont{{Pavan}}},
  \bibinfo{author}{\bibfnamefont{R.}~\bibnamefont{{Turolla}}},
  \bibinfo{author}{\bibfnamefont{S.}~\bibnamefont{{Zane}}}, \bibnamefont{and}
  \bibinfo{author}{\bibfnamefont{L.}~\bibnamefont{{Nobili}}},
  \bibinfo{journal}{\mnras} \textbf{\bibinfo{volume}{395}},
  \bibinfo{pages}{753} (\bibinfo{year}{2009}).

\end{thebibliography}

\end{document}